\documentclass[twocolumn,amssymb,nobibnotes,aps,prbstab, reprint,letterpaper,pre/printnumbers,amsmath,
superscriptaddress,%longbibliography,
showpacs,floatfix]{revtex4-1}
\usepackage{graphicx,xcolor}% Include figure files [dvips]
\usepackage{dcolumn}% Align table columns on decimal point
\usepackage{bm}% bold math
\usepackage{graphics}
\usepackage{float}
\usepackage{epsfig}
\usepackage{multirow}
\usepackage{amsmath}
\usepackage{multirow}
\usepackage{tabularx}
\usepackage{mathrsfs}
\usepackage{tabularx}
\usepackage{tabulary}

\usepackage[pass,paperwidth=8.5in,paperheight=11in]{geometry}
\usepackage[linktocpage=true]{hyperref}
\usepackage{hyperref}
\usepackage{comment}
\usepackage{soul}
\usepackage{notes2bib}
\hypersetup{
  pdfnewwindow=true, 
  colorlinks=true,
  linkcolor=blue, 
  anchorcolor=blue,
  citecolor=blue, 
  filecolor=blue,
  menucolor=blue, 
  urlcolor=blue}

\usepackage[normalem]{ulem}
\usepackage{xcolor}
\usepackage{orcidlink}

\begin{document}

\title{Exploring nonlinear Rashba effect and spin Hall conductivity in Janus MXenes W$_2$COX (X = S, Se, Te)}% Force line breaks with \\

\author{Arjyama Bordoloi\,\orcidlink{0009-0006-2760-3866}}
\affiliation{Department of Mechanical Engineering, University of Rochester, Rochester, New York 14627, USA}

\author{Sobhit Singh\,\orcidlink{0000-0002-5292-4235}}
\email{s.singh@rochester.edu}
\affiliation{Department of Mechanical Engineering, University of Rochester, Rochester, New York 14627, USA}
\affiliation{Materials Science Program, University of Rochester, Rochester, New York 14627, USA}

\date{\today}

\begin{abstract}

Rashba spin-orbit coupling (RSOC) facilitates spin manipulation without relying on an external magnetic field, opening up exciting possibilities for advanced spintronic devices. In this work, we examine the effects of crystal momentum ($k$) nonlinearity and anisotropy on the conventional Rashba effect, with a particular focus on their impact on the spin Hall conductivity (SHC) in a newly predicted family of 2D Janus materials, W$_2$COX (X = S, Se, Te). 
Using first-principles density functional theory calculations, we confirm the dynamical and mechanical stability of the studied 2D materials. Strikingly, this materials family exhibits pronounced nonlinear Rashba spin splitting at the $\Gamma$ point of Brillouin zone near the Fermi level, which cannot be adequately described by the linear-$k$ Rashba model. 
Therefore, third-order momentum contributions ({$k^3$}) must be incorporated into the Rashba Hamiltonian. 
Our analysis reveals that among the studied systems, W$_2$COS exhibits the highest {$k^3$} contribution of $-45.9$\,eV\,{\AA$^3$}, despite having the lowest linear Rashba constant. 
A detailed analysis of electronic structure reveals topological nontrivial behaviour in these 2D materials, yielding sizable SHC that is primarily governed by the nonlinear Rashba effect. Notably, these materials also exhibit large spin Hall angle (0.018 -- 2.5), which is comparable to that of in bulk topological insulators like Bi$_2$Se$_3$ and Bi$_2$Te$_3$, and surpassing those in narrow bandgap bulk semiconductors GeTe and SnTe, as well as heavy metals such as Pt. 
Sizable SHC, large spin Hall angles, and the ability to tune SHC via electric fields without altering the topological properties, rooted in the crystal field splitting, underscore the potential of these materials for spintronic applications.

\end{abstract}

%Use showkeys class option if keyword

\maketitle

%\tableofcontents

\section{Introduction}
Two-dimensional topological insulators (TIs), also known as quantum spin Hall (QSH) insulators, are characterized by an insulating bulk state combined with conducting helical edge states. These helical edge states enable counter-propagating motion of electrons with opposite spins, creating a ``two-lane highway" that allows for dissipationless electron transport ~\cite{Murakami_Science_2003, qi2011_rev_mod_phy, Tang2017_nature}. This unique property makes 2D TIs highly promising for low-power and multifunctional spintronic devices. Additionally, large band-gap 2D TIs are crucial for realizing topological superconductivity and Majorana fermions~\cite{Sau_PhysRevLett.104.040502,Sau_PhysRevB.82.214509}.

In recent years, there has been a surge in theoretical studies aimed at proposing efficient 2D TIs for spintronic applications, though only a few have been experimentally synthesized so far~\cite{Weng_MRS_Bulletin_2014,Ando_Journal_of_phy_soc_Japan, bernevig2006_science, konig2007_science, konig2008_JPPJ, ren2016_REPRTS_ON_PROGRESS_PHYSICS, SinghJPCL2019,Roche_2024}. In the quest for experimentally viable 2D TIs, Weng \textit{et al.} introduced a novel class of 2D materials known as M$_2$CO$_2$ (where M = Mo, W, Cr)~\cite{Weng_PhysRevB.92.075436}.This class is particularly intriguing due to its close association with the experimentally synthesized 2D MXenes, which are derived from the selective chemical etching of MAX phases—M$_{n+1}$AX$_n$ (n=1, 2, 3,...), where M is a transition metal, A is a group 12–14 element, and X is carbon (C) or nitrogen (N)~\cite{Barsoum_progree_in_solid_state_chem}. MXene surfaces are chemically active and commonly terminated with atoms or groups such as fluorine (F), oxygen (O), or hydroxyl (OH)~\cite{Harris_Jornal-of_phy_chem_c_2015, Khazaei_Adv_functional_mat,Khazaei_phys_chem_chem_phys_2014}. Therefore, M$_2$CO$_2$ holds promise for experimental realization, potentially advancing the development of 2D TIs. In these materials, band inversion arises from crystal field splitting, while the band gap is opened by spin-orbit coupling (SOC)~\cite{Weng_PhysRevB.92.075436,Si_Nano_Letters_2016,Yang_J_Mater_Chem_A_D2TA07161D,Yang_Phy_Chem_Chem_Phy_D3CP05142K, MAGHIRANG_chinese_journal_of_phy_20222346}. Among the M$_2$CO$_2$ monolayer family, W$_2$CO$_2$ stands out with the largest theoretically predicted band gap of 0.194 eV~\cite{Weng_PhysRevB.92.075436}, highlighting its potential applications in semiconductor spintronic technology. 

However, for a material to be suitable for spintronic device design, it is crucial that its electronic properties can be tuned via external perturbations, such as electric fields or strain, to enhance device functionality. Unlike intrinsic SOC, as found in W$_2$CO$_2$ monolayer, which lacks such tunability, Rashba spin-orbit coupling (RSOC) offers greater flexibility, making it more suitable for device applications ~\cite{heide2006spin, Bihlmayer_2015, Premasiri_2019, Barla2021}. Therefore, in this work, we aim to induce RSOC in centrosymmetric W$_2$CO$_2$ by breaking its spatial-inversion symmetry through the substitution of one oxygen atom with a chalcogen atom X (X = S, Se, Te). This results in a new series of 2D Rashba materials, W$_2$COX. The Rashba effect, arising from the broken inversion symmetry, introduces an additional degree of freedom, thereby enhancing the potential of these materials for spintronic devices with more flexible and tailored design possibilities.
%However, for a material to be suitable for the design of spintronic devices, the ability to tune its electronic properties using external perturbations such as electric fields or strain is crucial. This tunability enhances device functionality by adding an extra degree of flexibility. One effective way to introduce such tunability in spintronic devices is through Rashba spin-orbit coupling (RSOC), which offers an advantage over intrinsic SOC, such as that present in W$_2$CO$_2$. which enhances device functionalities by adding a additional degree of flexibility. One such way to add tunability to spintronic devices is Rashba spin-orbit coupling (RSOC) which offers advantage over intrinsic SOC aas is present in W$_2$CO$_2$.
%Intrinsic SOC, as found in W$_2$CO$_2$, limits this tunability. In contrast, Rashba spin-orbit coupling (RSOC) offers greater control over electronic properties because it can be easily manipulated using external gate voltages or strain. 

In this work, we start with the most stable configuration of W$_2$CO$_2$ and substitute one of the oxygen atoms with a similar chalcogen atom X (S, Se, Te). Using first-principles methods, we investigate how these substitutions influence the electronic and topological properties compared to the parent compound. Our findings reveal that the substitution breaks inversion symmetry, leading to noticeable Rashba spin splitting in each material.  Interestingly, due to the C$_{3v}$ point group symmetry and deviations from a purely parabolic band structure, the standard linear in crystal momentum ($k$) Rashba-Bychkov model alone cannot adequately explain the observed Rashba spin splitting in these materials. Therefore, third-order momentum contributions ($k^3$)  need to be incorporated into the Rashba Hamiltonian. We observe that all three monolayers show significant nonlinear RSOC, with W$_2$COS exhibiting the highest $k^3$ contribution (-45.9 eV\AA$^3$) in its conduction band near the Fermi level (E$_F$). 
In the context of topological properties, while W$_2$COS and W$_2$COSe monolayers remain topologically nontrivial ($Z_2 = 1$) after the substitution, W$_2$COTe monolayer becomes topologically trivial ($Z_2 = 0$).

Further, we examine the spin Hall conductivity (SHC) for these monolayers, which is crucial given their non-trivial topological features and significant RSOC, making them promising candidates for achieving sizable SHC. We find W$_2$COS monolayer exhibits the highest transverse spin Hall conductivity, approximately 474 (\(\hbar/e) \text{S/cm}^{-1}\) near E$_F$, attributed to dominant $k^3$ contributions. Interestingly, despite showing a lower SHC compared to heavier metals like platinum~\cite{Guo_PhysRevLett.100.096401} and tantalum~\cite{Sagasta_PhysRevB.98.060410}, W$_2$COX exhibits relatively larger spin Hall angles (0.018--2.5 at E$_F$), highlighting their potential for designing efficient spintronic devices. Additionally, since the band inversion in these materials is primarily due to crystal field splitting, it is also expected that the SHC can potentially be modulated by applying electric fields, without altering their topological electronic properties.

\begin{figure}[!!b]
\centering
\includegraphics[width=1\columnwidth]{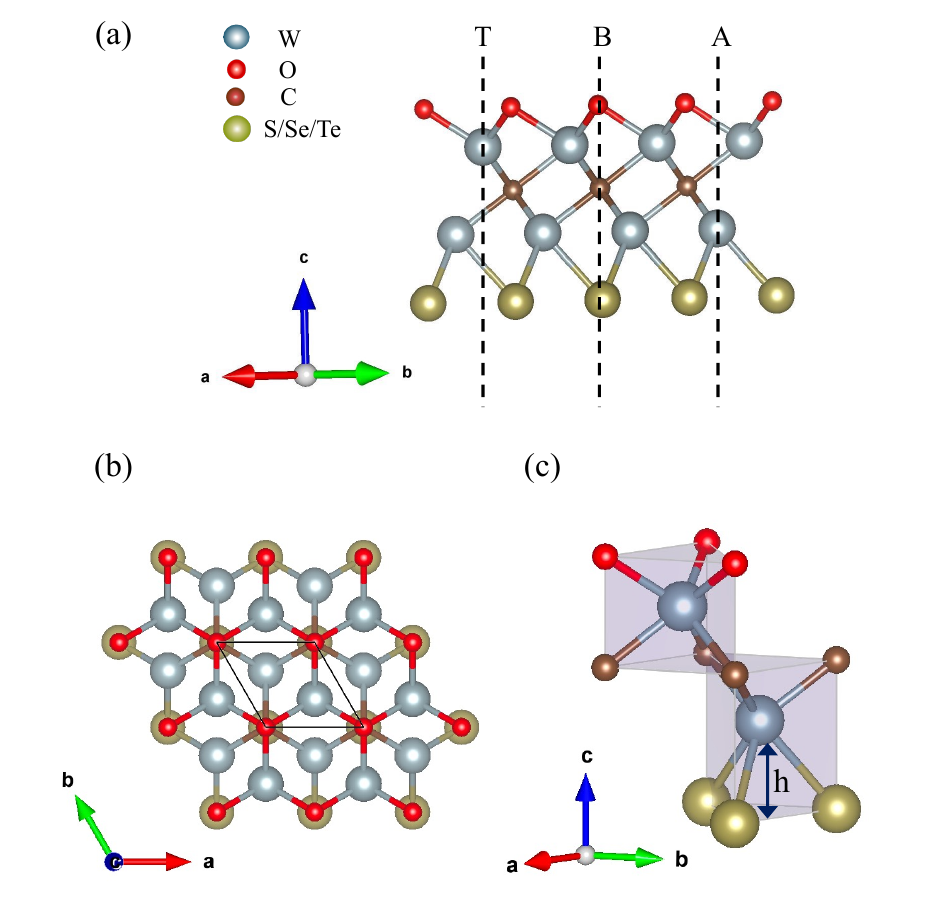}
\caption{Optimized crystal structure of W$_2$COX (X=S, Se, Te) monolayers generated using VESTA~\cite{VESTA}. (a) represents the side view, (b) represents the top view and (c) represents the crystal coordination environment in the triangular prism crystal field configuration. }
\label{fig: structure}
\end{figure}

\section{Computational details}

First principles density functional theory (DFT) calculations \cite{HK_dft_1964, KS_dft_1965} were performed using the Vienna Ab initio Simulation Package (VASP) \cite{Kresse96a, Kresse96b, KressePAW} on free-standing monolayers of W$_2$COX (X=S, Se, Te). A vacuum thickness of 15~\AA~was maintained along the z-direction to avoid interlayer interactions. The projector augmented-wave (PAW) method \cite{Blochl94} was employed with a kinetic energy cutoff set at 650 eV for the plane wave basis set. The exchange-correlation functional was treated within the generalized gradient approximation as parameterized by Perdew, Burke, and Ernzerhof for solids (PBEsol) \cite{PBEsol}. All the lattice parameters and inner atomic coordinates were fully optimized until the ground state was reached, with an electronic energy convergence criterion of 10$^{-7}$ eV and residual Hellmann-Feynman forces of less than 10$^{-3}$ eV/Å per atom. A $\Gamma$-centered k-mesh of size 12\,$\times$\,12\,$\times$\,1 was used for both geometrical optimization and electronic structure calculations. Contributions from six valence electrons each from W\,($6s^25d^4$), O\,($2s^22p^4$), and S\,($3s^23p^4$)/Se\,($4s^24p^4$)/Te\,($5s^25p^4$) were considered in the PAW pseudopotential, while four electrons were considered for C\,($2s^22p^2$).

\begin{figure*}
\centering
\includegraphics [trim=0.0cm 1cm 0cm 0cm, clip=true,scale=0.42, width=18 cm]
{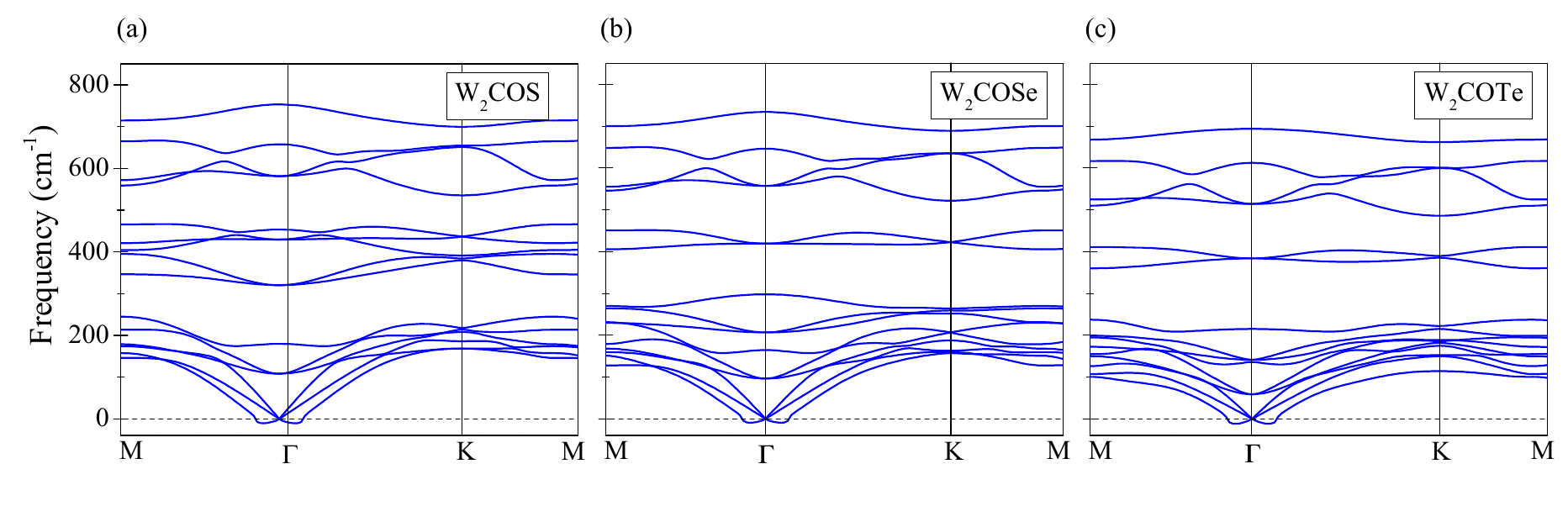}
\caption{Phonon band structure of (a) W$_2$COS, (b)W$_2$COSe, and (c)W$_2$COTe monolayers}
\label{fig:phonon}
\end{figure*}

The dynamical stability of the studied materials was tested using phonon calculations in 4\,$\times$\,4\,$\times$\,1 supercells within the finite displacement method in VASP. The post-processing of phonon calculations was performed using PHONOPY \cite{phonopy}. The elastic constants C$_{ij}$ were computed using the stress-strain relationship as implemented in VASP, ensuring the convergence of the elastic constants with varying k-mesh sizes. The {\sc MechElastic} Python package \cite{MechElastic,Mechelastic_Comp_Phys_comm_2021} was used for detailed analysis of all elastic constants and for checking the mechanical stability of the systems. SOC was considered in all calculations except the phonon calculations. VASPKIT~\cite{VASPKIT} and the {\sc PyProcar} package \cite{pyprocar} was used for post-processing of the electronic structure data. 

Topological properties were investigated using Wannier tight-binding Hamiltonians with projections from the s- and d-orbitals of W, and the s- and p-orbitals of C, O, and Te atoms, computed using Wannier90 \cite{Marzari2012}. Subsequent calculations of Z$_2$ topological invariants from the Wannier tight-binding Hamiltonian were performed using WannierTools \cite{WU2018405}. SHC for each of the investigated systems was also computed from the Maximally Localized Wannier Functions (MLWF), using a dense k-mesh of 401\,$\times$\,401\,$\times$\,1 as implemented in WannierTools \cite{WU2018405}. The SHC values were converged to within less than 1\% by varying the k-mesh size.

\section{Results and discussions}

\subsection{Crystal structure and dynamical and mechanical stability}\label{secA}

Figure~\ref{fig: structure} illustrates the crystal structure of W$_2$COX (where X = S, Se, or Te) monolayer. These compounds are derived from the parent W$_2$CO$_2$ structure by substituting one oxygen atom with a chalcogen atom X (X = S, Se, Te). Previous studies~\cite{Weng_PhysRevB.92.075436,Si_Nano_Letters_2016,Yang_J_Mater_Chem_A_D2TA07161D,Yang_Phy_Chem_Chem_Phy_D3CP05142K,MAGHIRANG_chinese_journal_of_phy_20222346} have shown that, among all possible configurations resulting from the termination of the transition metal atom site in W$_2$C with oxygen atoms, the BB-terminated configuration is energetically the most favorable one. 
Therefore, 
%to achieve our goal of studying the effect of broken inversion symmetry on the electronic and topological properties compared to the parent compound W$_2$CO$_2$, 
we start from the most stable BB-terminated configuration and substitute one oxygen atom with a chalcogen atom (X) at the B-site, as shown in Fig.~\ref{fig: structure}(a), to break the spatial-inversion symmetry. Our investigation reveals that due to this substitution, all three systems—W$_2$COX (X = S, Se, Te)—crystallize with non-centrosymmetric $p3m1$ symmetry (layer group number 69), in contrast to W$_2$CO$_2$, which crystallizes in centrosymmetric $p\bar{3}m1$ symmetry (layer group number 72). This is consistent with some of the the previously reported literature on similar type of systems including Mo$_2$COX (X=S, Se, Te)~\cite{Karmakar_PhysRevB.107.075403}. The optimized lattice parameters and relevant bond lengths for each of these systems are provided in Table \ref{tab:crystal parameters} in Appendix \ref{Appendix A}. 

\begin{figure*}
\centering
\includegraphics[trim=0.0cm 0cm 9cm 0cm, clip=true,scale=0.42, width=18 cm]{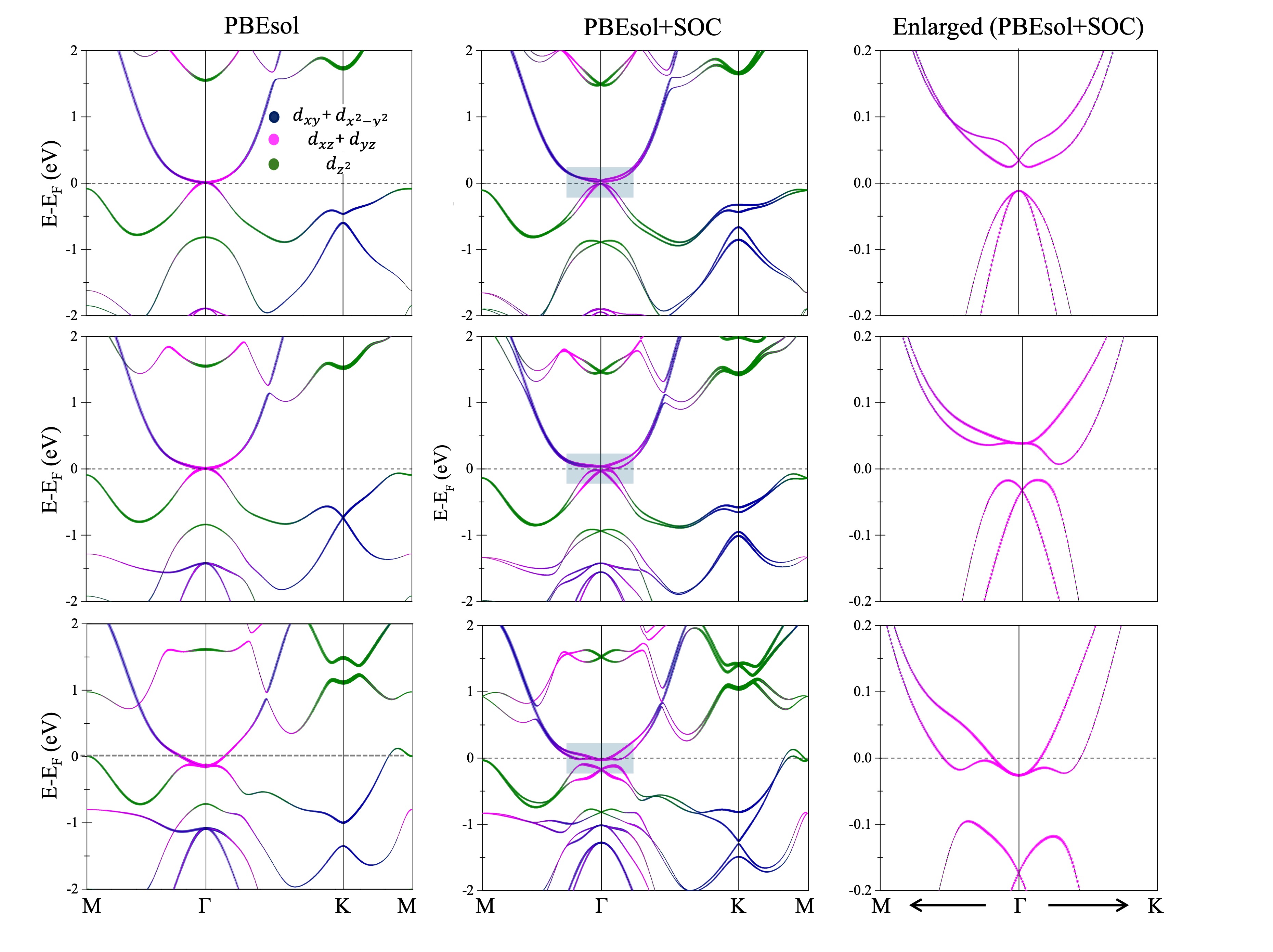}
\caption{Electronic band structures of W$_2$COX (X = S, Se, Te) monolayers with projections from the \(d_{x^2-y^2} + d_{xy}\), \(d_{yz} + d_{xz}\), and \(d_{z^2}\) orbitals of W (5d). The left panel shows the electronic band structure without SOC using PBEsol, the middle panel includes SOC (PBEsol+SOC), and the right panel provides an enlarged view of the shaded portion of the electronic band structure calculated with PBEsol+SOC near the Fermi level around the $\Gamma$ point.}
\label{fig: electronic band structures}
\end{figure*}

For any theoretically predicted material, it is crucial to investigate its dynamical and mechanical stability. Figures \ref{fig:phonon}(a), \ref{fig:phonon}(b), and \ref{fig:phonon}(c) represent the phonon band structure of W$_2$COS, W$_2$COSe, and W$_2$COTe monolayers, respectively, computed along the high-symmetry directions in the Brillouin zone. No imaginary phonon modes are present in the phonon band structure, except for small U-shaped features observed in the first acoustic branch near the $\Gamma$ point, which is a characteristic of layered 2D systems \cite{Stengel_PhysRevX.11.041027, SinghPRB2017, SinghPRL2020}. This verifies the dynamical stability of all three monolayers. 
%It's important to note that the small U-shaped imaginary features near the $\Gamma$ point do not indicate lattice instability; rather, they correspond to flexural acoustic modes 
%These features could ideally be minimized or eliminated by using a much denser q grid or by introducing higher order terms\cite{Stengel_PhysRevX.11.041027}. As anticipated, the frequency of the highest phonon mode decreases from X=S to Se to Te, which can be attributed to the increase in atomic mass with increasing atomic number.

Next, we ensure the mechanical stability of these materials. We compute the elastic constants \(C_{ij}\) for each material and check if they satisfy the Born-Huang mechanical stability criteria~\cite{Born_Huang,nye1985physical}. According to these criteria a crystal is mechanically stable if it has the lowest Gibbs free energy in its relaxed state, i.e., in the absence of external loads, compared to any other state obtained by applying a small strain. This means the elastic stiffness matrix, \(C_{ij}\), must be positive definite, with all eigenvalues positive, and the matrix must be symmetric. Our calculations reveal that all the three studied systems meet the necessary conditions for mechanical stability, with positive values satisfying the Born-Huang mechanical stability criteria for 2D hexagonal structures, i.e., \(C_{11} + C_{12} > 0\) and \(C_{11} - C_{12} > 0\). We also calculate the mechanical constants including Young's modulus, 2D layer modulus and shear modulus which are listed in table \ref{tab:crystal parameters} in Appendix \ref{Appendix A}. The values are comparable to those of other standard 2D materials like graphene and h-BN~\cite{Andrew_PhysRevB.85.125428}.
%The values of \(C_{11}\) (\(C_{12}\)) are 381.8 N/m (121.4 N/m), 374.9 N/m (104.3 N/m), and 327.4 N/m (81.1 N/m) for W$_2$COS, W$_2$COSe, and W$_2$COTe, respectively. These elastic constants indicate that all three systems meet the necessary conditions for mechanical stability, with positive values satisfying the Born-Huang mechanical stability criteria for 2D hexagonal structures, i.e., \(C_{11} + C_{12} > 0\) and \(C_{11} - C_{12} > 0\). We also calculated the mechanical constants, including Young's modulus (343.4, 345.9, and 307.8 N/m for W$_2$COS, W$_2$COSe, and W$_2$COTe, respectively), 2D layer modulus (252.3, 239.8, and 204.4 N/m for W$_2$COS, W$_2$COSe, and W$_2$COTe, respectively), and shear modulus (131.4, 136.6, and 123.1 N/m for W$_2$COS, W$_2$COSe, and W$_2$COTe, respectively). These values are comparable to those of other standard 2D materials like graphene and h-BN~\cite{Andrew_PhysRevB.85.125428}.

\subsection{Electronic band structure}\label{secB}

Figure \ref{fig: electronic band structures} illustrates the electronic band structures of W$_2$COX (X=S, Se, Te) monolayers with projections from the d$_{x^2-y^2}$+d$_{xy}$, d$_{xz}$+d$_{yz}$, and d$_{z^2}$ orbitals of W. Notably, the W-5d orbitals are under the influence of a triangular prismatic crystal field of the C$_{3v}$ point group, as shown in Fig.~\ref{fig: structure}(c). The band structures are computed along the high symmetry directions of the Brillouin zone with the dashed line representing the Fermi level. 
%The left panel shows the electronic band structure without SOC (PBEsol), while the middle panel includes SOC (PBEsol+SOC). 
Without SOC, all three systems exhibit semimetallic behavior. In W$_2$COS and W$_2$COSe, the valence band maximum (VBM) and conduction band minimum (CBM) touch each other at the \(\Gamma\) point at the Fermi level (E$_F$). However, in W$_2$COTe, the CBM forms an electron pocket at the $\Gamma$ point, and the VBM forms a hole pocket near M point of the Brillouin zone. 
%In the cases of W$_2$COS and W$_2$COSe, the VBM and CBM touch each other at the Fermi level. However, for W$_2$COTe, the CBM forms an electron pocket exactly at the $\Gamma$ point and the VBM forms a hole pocket at the M point along the K-M direction. 

Focusing specifically on the orbital contributions near the Fermi level, the primary contributions come from the W-5d orbitals.
%d$_{xz}$+d$_{yz}$ and d$_{z^2}$ orbitals. This is distinct from W$_2$CO$_2$, where the main contribution arises from the W d$_{x^2-y^2}$+d$_{xy}$ and d$_{z^2}$ orbitals~\cite{Weng_PhysRevB.92.075436,Si_Nano_Letters_2016,Yang_J_Mater_Chem_A_D2TA07161D,Yang_Phy_Chem_Chem_Phy_D3CP05142K, MAGHIRANG_chinese_journal_of_phy_20222346}. 
Even without SOC, the orbital-projected band structure indicates a band inversion between the d$_{xz}$+d$_{yz}$ and d$_{z^2}$ orbitals for all three materials. Furthermore, as we move from S to Se to Te, the bands originating from the d$_{xz}$+d$_{yz}$ and d$_{x^2-y^2}$+d$_{xy}$ orbitals are pushed up in energy near the $\Gamma$ point. This trend is primarily due to the decrease in electronegativity from S to Se to Te with increasing atomic number. In contrast, the d$_{z^2}$ orbitals are not significantly affected, largely because they do not interact directly with the ligands.

The inclusion of SOC induces spin splitting in the electronic band structures. Due to the broken inversion symmetry, W$_2$COX monolayers exhibit a Rashba-type spin splitting at the \(\Gamma\) point, which was absent in the parent compound W$_2$CO$_2$.
%W$_2$COSe and W$_2$COTe exhibit a Rashba-type spin splitting at the \(\Gamma\) point in the VBM, whereas in W$_2$COS, the Rashba effect is evident in the CBM, which was absent in the parent compound W$_2$CO$_2$. 
For practical spintronic device design, it is crucial to accurately determine the Rashba parameter of the material, as it provides a direct estimation of the spin precession length essential for device design. For the investigated systems, if we focus on the enlarged view of the Rashba bands (Figure 2) near the \(\Gamma\) point, the bands exhibit a non-quadratic behaviour. While most of the recent literature often overlooks this feature and uses the linear (in crystal momentum k) Rashba model to fit such bands, there are studies emphasizing the necessity of including higher-order terms in the Rashba Hamiltonian for systems with certain high symmetries~\cite{Vajna_PhysRevB.85.075404,Yang_PhysRevB.74.193314,Gupta_JACS_2021}. This is crucial for accurately capturing spin splitting and avoiding underestimation or overestimation of the Rashba coupling strength. 

\begin{figure}[!!b]
\centering
\includegraphics[trim=0.2cm 0cm 2.5cm 1cm, clip=true,scale=0.42, width=1\columnwidth]{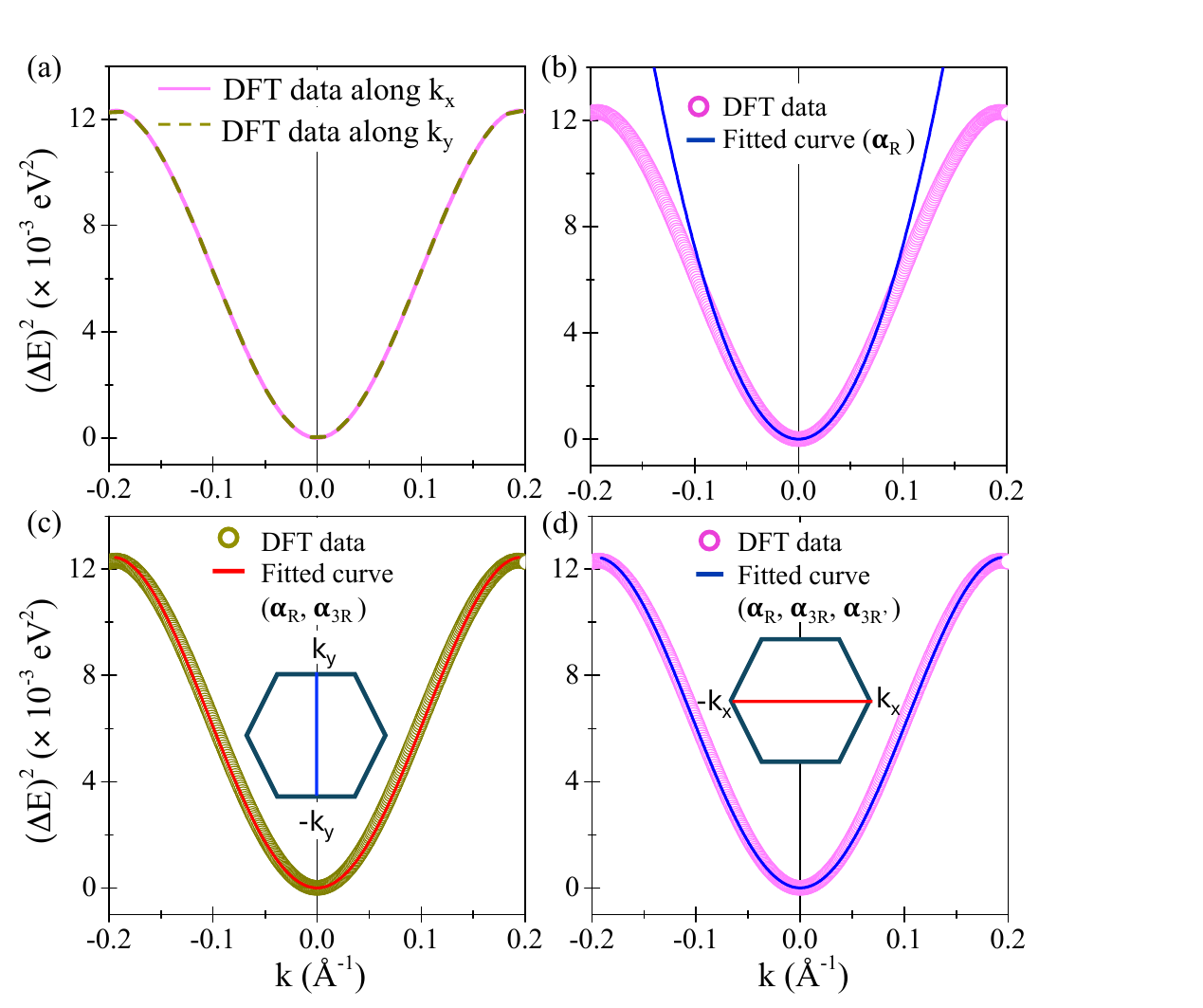}
\caption{(a) Square of the energy splitting $\Delta E = [E_+(k) - E_-(k)] / 2$ of Rashba bands in W\(_2\)COTe as a function of \(k\). The magenta line represents data along the $k_x$ direction, while the dark yellow dashed line represents data along the $k_y$ direction. (b)-(d) Fitting of DFT data to different functions to obtain first-order and third-order Rashba parameters. Solid lines indicate the fitted curves, while circular patterns represent the DFT data.}
\label{fig: k3 fitting}
\end{figure}

\begin{figure*}
\centering
\includegraphics[width=18 cm]{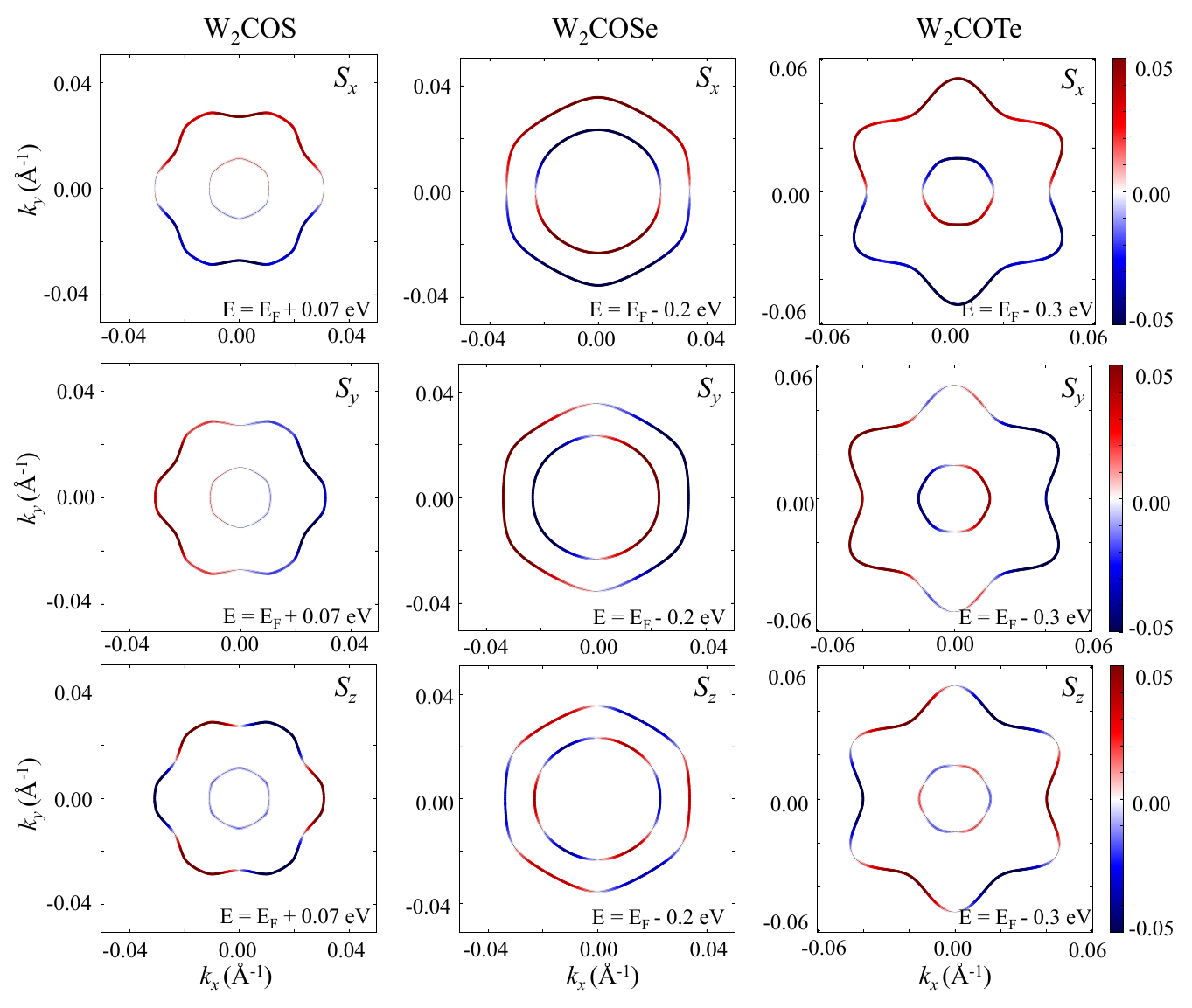}
\caption{Constant energy 2D contour plots of spin texture calculated for W\(_2\)COX (X = S, Se, Te) monolayers in the \(k_x\)-\(k_y\) plane, centered at the \(\Gamma\) point near the Fermi level (E$_F$).}
\label{fig: spin texture}
\end{figure*}

As reported in literature~\cite{Vajna_PhysRevB.85.075404}, the Rashba spin splitting in materials with C$_{3v}$ point group symmetry, which have non-parabolic bands, cannot be adequately described by the linear-$k$ Rashba-Bychkov model alone. To accurately capture the Rashba spin splitting present in these materials, terms containing cubic orders of \(k\) need to be incorporated into the Rashba Hamiltonian. Hence, in the case of C$_{3v}$ symmetry, the Rashba Hamiltonian takes the form
\begin{equation}
\label{hamiltonian}
    \begin{aligned}
        \hat{H}_{R}(k) &= (\alpha_{R}k + \alpha_{3R}k^3)(\cos\phi \sigma_{y} - \sin\phi \sigma_{x}) \\
        &\quad + \alpha_{3R}'^2 k^3 \cos 3\phi \sigma_{z}\,,
    \end{aligned}
\end{equation}
where $\phi = \cos^{-1}(k_x/k)$ with $k = \sqrt{k_x^2 + k_y^2}$. 
The $k_x$ and $k_y$ directions are denoted in the inset of Figs.~\ref{fig: k3 fitting}(c,d). 
%The \textit{x}-direction is considered from $-k_x$ to $k_x$ of the hexagonal Brillouin zone as shown in Fig.~\ref{fig: k3 fitting}(d). 
Furthermore, $\alpha_{R}$ represents the linear Rashba contribution, $\alpha_{3R}$ represents the isotropic third-order Rashba contribution, and $\alpha_{3R}'$ is the anisotropic third-order Rashba contribution. 
The energy eigenvalues of the full Hamiltonian for 2D free electron gas including Rashba term %$\hat{H}_{R}(k)$
\begin{equation}
    \begin{aligned}
     \hat{H}(k) =  \hat{H}_{0}(k) +  \hat{H}_{R}(k) \equiv \frac{k^2}{2m} +  \hat{H}_{R}(k) 
    \end{aligned}
\end{equation}
are 
\begin{equation}
    \begin{aligned}
        E_\pm(k) &= \frac{k^2}{2m}\pm \sqrt{{(\alpha_{R}k + \alpha_{3R}k^3)}^2 + \alpha_{3R}'^2 k^6 {\cos}^2 3\phi}\,.
    \end{aligned}
\end{equation}

To determine the accurate values of these Rashba terms, we adopt a similar approach to that used by Vajna \textit{et al.}~\cite{Vajna_PhysRevB.85.075404}. First, we compute the square of the energy splitting of the two eigenvalues of the Rashba Hamiltonian, i.e., $\Delta E = [E_+(k) - E_-(k)] / 2$. The square of this energy difference attains the form
\begin{equation}
    \begin{aligned}
        {\Delta E(k)}^2 = {(\alpha_{R}k + \alpha_{3R}k^3)}^2 + \alpha_{3R}'^2 k^6 {\cos}^2 3\phi\,.
    \end{aligned}
\end{equation}

Figure \ref{fig: k3 fitting}(a) represents the variation of \({\Delta E(k)}^2\) with respect to \(k\), plotted along both the \(k_x\) and \(k_y\) directions for the representative case of W$_2$COTe monolayer. Similar plots for W$_2$COS and W$_2$COSe monolayers are included in the Appendix \ref{Appendix B}. The magenta line represents the data plotted along the \(x\) direction, while the dark yellow dashed line represents the data plotted along the \(y\) direction. For all three materials, the variations of \({\Delta E(k)}^2\) along the \(k_x\) and \(k_y\) directions overlap on top of each other, indicating negligible anisotropy in the electronic band structure between the \(k_x\) and \(k_y\) directions. This observation aligns well with our results of the anisotropic third-order Rashba term.

We calculate the linear-$k$ Rashba term \(\alpha_{R}\) by fitting the linear Rashba-Bychkov model to the DFT-computed data, as shown in Fig.~\ref{fig: k3 fitting}(b). As can be seen from Figs.~\ref{fig: k3 fitting}(b) and  \ref{fig: W2COS/Se_higher_order}(b), the linear Rashba model fits well with the DFT-computed results only for \(\lvert k_x \rvert \leq 0.03 \, \text{Å}^{-1}\), \(0.09 \, \text{Å}^{-1}\), and \(0.07 \, \text{Å}^{-1}\) for X = S, Se, and Te, respectively. Beyond these values, higher-order terms are required to accurately fit the DFT-computed bands. The fitting yields \(\alpha_{R}\) values of 0.45, 0.59, and 0.85 eVÅ for X = S, Se, and Te, respectively.

Next, we determine the isotropic cubic contribution \(\alpha_{3R}\) by fitting the DFT data along the \(k_y\) direction, setting \(k_x = 0\) and keeping \(\alpha_{R}\) constant as obtained from the linear Rashba model [see Figure \ref{fig: k3 fitting}(c)]. In this case, \(\phi = \pi/2\). The fitting yields \(\alpha_{3R}\) values of -45.9, -3.7, and -7.3 eVÅ\(^3\) for X = S, Se, and Te, respectively. Strikingly, the CBM of W$_2$COS has the highest cubic contributions, which closely correlates with our spin Hall conductivity results discussed below.

Finally, we compute the anisotropic contribution by fitting the DFT data along the \(k_x\) direction with \(k_y = 0\), resulting in \(\phi =0\) [see Figure \ref{fig: k3 fitting}(d)]. We keep the values of \(\alpha_{R}\) and \(\alpha_{3R}\) fixed at the previously obtained values and observe that the anisotropic contributions are nearly zero, {\it i.e.}, on the order of \(10^{-4}\). This agrees well with our observations in Figure \ref{fig: k3 fitting}(a), which shows the absence of anisotropy in electronic band structure.

Another significant outcome of including SOC in the electronic band structure is the emergence of a gap between the VBM and CBM in all three cases. With the opening of this band gap, W$_2$COS and W$_2$COSe monolayers acquire a semiconducting nature, each with an indirect band gap of 36 meV and 24 meV, respectively. In contrast, W$_2$COTe retains its semimetallic nature even with SOC included, due to the presence of the electron and hole pockets. 
%To eliminate the possibility of underestimating the bandgap with the PBEsol functional, we also performed band structure calculations using the Hybrid functional HSE06~\cite{HSE06}. We observed that while the bandgap remains zero without SOC, it increases to 179 meV for W$_2$COS and to xx meV for W$_2$COSe when SOC is included. 
While the opening of this band gap along with the band inversion observed in the electronic band structure are indicative of non-trivial topology, confirmation through the calculation of the \( Z_2 \) topological invariant is essential, which is discussed in section \ref{secC}.

Figure~\ref{fig: spin texture} shows constant energy 2D contour plots of spin texture calculated for all three systems in the \( k_x \)-\( k_y \) plane centered at the \(\Gamma\) point. The helical spin texture, typical of 2D Rashba systems, confirms the presence of the 2D Rashba effect in these materials. Strikingly, all three systems exhibit contributions from the \(S_z\) component along with \(S_x\) and \(S_y\) components. This suggests the presence of cubic terms of \(k\) in the Rashba Hamiltonian, leading to a hexagonal warping effect in the spin texture. The warping effect is relatively less pronounced in W$_2$COSe due to smaller values of the cubic Rashba term \(\alpha_{3R}\), whereas it becomes more prominent in the other two systems with an increase in the value of \(\alpha_{3R}\).

\subsection{Topological properties}\label{secC}
To analyze the topological properties of the studied configurations, we employ the Z$_2$ topological invariant calculated using the concept of Wannier charge centers (WCC)~\cite{Soluyanov_PhysRevB.83.035108}. The evolution of the WCCs of maximally localized Wannier functions along the time-reversal invariant plane k$_z$ = 0 is used to characterize their topological nature. Using WannierTools, we compute the Z$_2$ topological invariants for all three systems. Interestingly, W$_2$COS and W$_2$COSe exhibit topologically nontrivial behavior with Z$_2$ = 1, whereas W$_2$COTe exhibits topologically trivial behavior with Z$_2$ = 0. However, the parent compound W$_2$CO$_2$ shows topologically nontrivial behavior~\cite{Weng_PhysRevB.92.075436,Si_Nano_Letters_2016,Yang_J_Mater_Chem_A_D2TA07161D,Yang_Phy_Chem_Chem_Phy_D3CP05142K,MAGHIRANG_chinese_journal_of_phy_20222346}.

To understand the origin of the topological behavior of these materials, we analyze the energy band diagrams of these systems near the Fermi level. From the orbital-projected band structures of these monolayers, it is evident that W-$5d$ orbitals dominate near the Fermi level in all three materials. These W-$5d$ orbitals are under the influence of a triangular prismatic crystal field of the C$_{3v}$ point group, as shown in Fig. \ref{fig: structure}(c). This crystal field splitting causes the doubly degenerate d$_{xz}$+d$_{yz}$ orbitals to be higher in energy than the d$_{z^2}$ orbitals, with the Fermi level lying between these two energy levels. The d$_{x^2-y^2}$+d$_{xy}$ orbitals, which are also doubly degenerate, on the other hand, are located well below the Fermi level. Hence, while analyzing the topological properties of these materials, we particularly focus on the d$_{xz}$+d$_{yz}$ and d$_{z^2}$ orbitals.

\begin{figure}[!!t]
\centering
\includegraphics[width=8.5 cm]{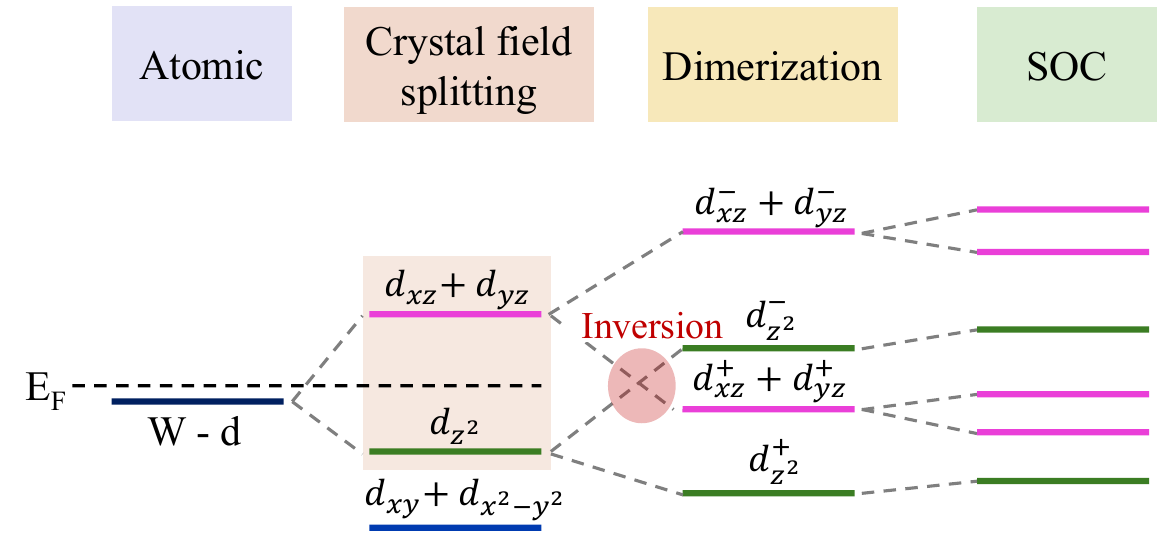}
\caption{Schematic representation of the electronic band evolution of W $5d$ orbitals at the $\Gamma$ point in W$_2$COX (X = S, Se, Te) monolayers.}
\label{fig: CFT}
\end{figure}

Since two W atoms are present in these systems, the W d-orbitals form bonding (+) and antibonding (-) states. As shown in Figure~\ref{fig: CFT}, a band inversion occurs between the antibonding d$_{z^2}$ orbitals and the bonding d$_{xz}$+d$_{yz}$ orbitals, leading to the topologically nontrivial nature of W$_2$COS and W$_2$COSe. The further inclusion of SOC introduces a spin degree of freedom, which opens a band gap around the Fermi level, as depicted in the band structure plots (Figure \ref{fig: electronic band structures}).

However, despite W$_2$COTe having a similar orbital-projected band structure to W$_2$COS and W$_2$COSe, it exhibits topologically trivial behavior and a semimetallic nature. This is likely due to the relatively stronger SOC in W$_2$COTe compared to the other two materials. The competitive interplay between SOC and crystal field splitting in W$_2$COTe probably leads to the loss of its topologically nontrivial characteristics~\cite{Weng_PhysRevB.92.075436,Si_Nano_Letters_2016,Yang_J_Mater_Chem_A_D2TA07161D,Yang_Phy_Chem_Chem_Phy_D3CP05142K,MAGHIRANG_chinese_journal_of_phy_20222346}. .

%\AB{Reason behind this!! crystal field splitting diagram}

\subsection{Spin Hall Conductivity}\label{secD}
Next, we compute the spin Hall conductivity (SHC) for studied monolayers. The spin Hall effect refers to the generation of a transverse pure spin current, perpendicular to the plane of charge and spin currents, induced by a longitudinal charge current~\cite{Hirsch_PhysRevLett.83.1834, DYAKONOV_phy_let_a_1971459, Jungwirth_Nature_Mat_2012}. This phenomenon is crucial for leveraging electron spins in information processing, a key objective in semiconductor spintronics. Rashba materials are particularly significant in this context due to their potential for electrically and non-volatilely controlling spins, enabling integration of storage, memory, and computing functionalities. Additionally, compared to heavy metals like platinum and tantalum, commonly used as spin-current generators, topological insulators are expected to consume less energy due to minimal electron backscattering, while achieving comparable spin Hall angles~\cite{Gong_PhysRevB.109.045124, Farzaneh_PhysRevMaterials.4.114202}. In this context, W$_2$COX (X = S, Se) monolayers are particularly significant because they not only possess non-trivial topological features but also exhibit significant RSOC, making them promising candidates for achieving sizable SHC. 

\begin{figure}[!!b]
\centering
\includegraphics[width=9 cm]{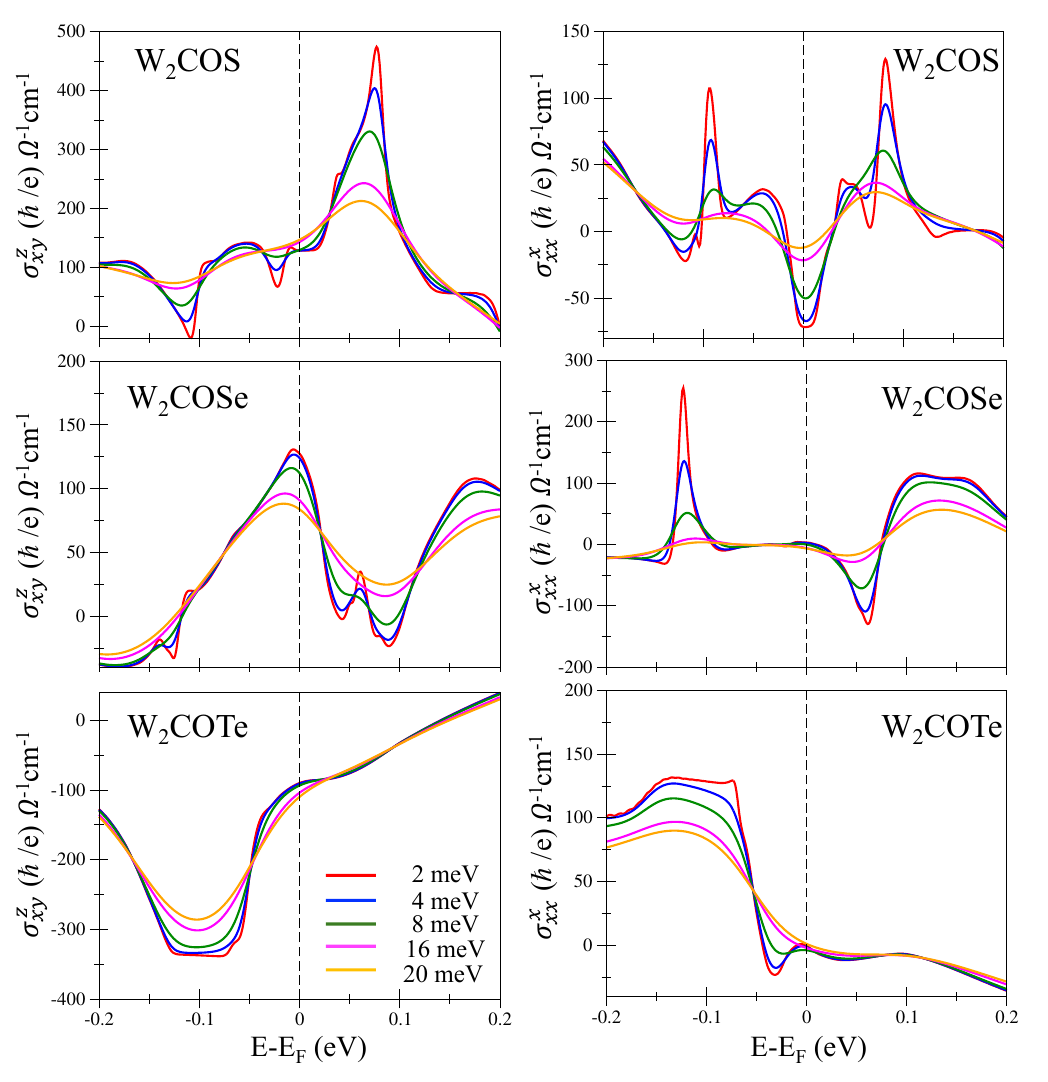}
\caption{Spin Hall conductivity in units of \((\hbar/e)\text{S/cm}^{-1}\) for W$_2$COS, W$_2$COSe, and W$_2$COTe monolayers as a function of chemical potential, with the broadening parameter \(\sigma\) varying from 2 meV to 20 meV. The left panel illustrates the variation of transverse spin conductivity \(\sigma_{xy}^z\), while the right panel shows the variation of longitudinal spin conductivity (\(\sigma_{xx}^x\)). The dashed vertical line represents the Fermi level.}
\label{fig: SHC}
\end{figure} 

According to linear response theory, the SHC, denoted as \({\sigma_{ij}^k}\), is a tensor that relates the electric field \(E_j\) to a spin current density \({J_{i}^k}\), expressed as \({J_{i}^k} = {\sigma_{ij}^k} E_j\). The SHC is a third-rank tensor, with indices \(i\) and \(j\) representing perpendicular directions. Although, in general, the SHC tensor should have 27 components, symmetry constraints reduce the number of non-zero elements based on selection rules. For the \(p3m1\) layer group, as determined using the Bilbao Crystallographic Server~\cite{Bilbao}, the SHC tensor has only six non-zero components: \(\sigma_{xx}^x\), \(\sigma_{yy}^x\), \(\sigma_{xy}^y\), \(\sigma_{yx}^y\), \(\sigma_{xy}^z\), and \(\sigma_{yx}^z\). Among these, only two are independent as the components are interrelated as \(\sigma_{xx}^x = -\sigma_{yy}^x = -\sigma_{xy}^y = -\sigma_{yx}^y\) and \(\sigma_{xy}^z = -\sigma_{yx}^z\). 
%We used WannierTools, which is based on the maximally localized Wannier functions, to calculate the SHC tensor values. 
Figure~\ref{fig: SHC} shows the SHC values in units of (\(\hbar/e)\text{S/cm}^{-1}\) for all three systems as a function of chemical potential near E$_F$, with the broadening parameter \(\sigma\) varying from 2 meV to 20 meV. However, lower broadening parameters in the range of 2-8 meV are anticipated to be more reasonable for these systems considering their small bandgap values. 
%The left panel illustrates the variation of \(\sigma_{xy}^z\), while the right panel shows the variation of \(\sigma_{xx}^x\). 

\begin{figure}[!!t]
\centering
\includegraphics[trim=0.2cm 0cm 0cm 2cm, clip=true,scale=0.42, width=9.5 cm]{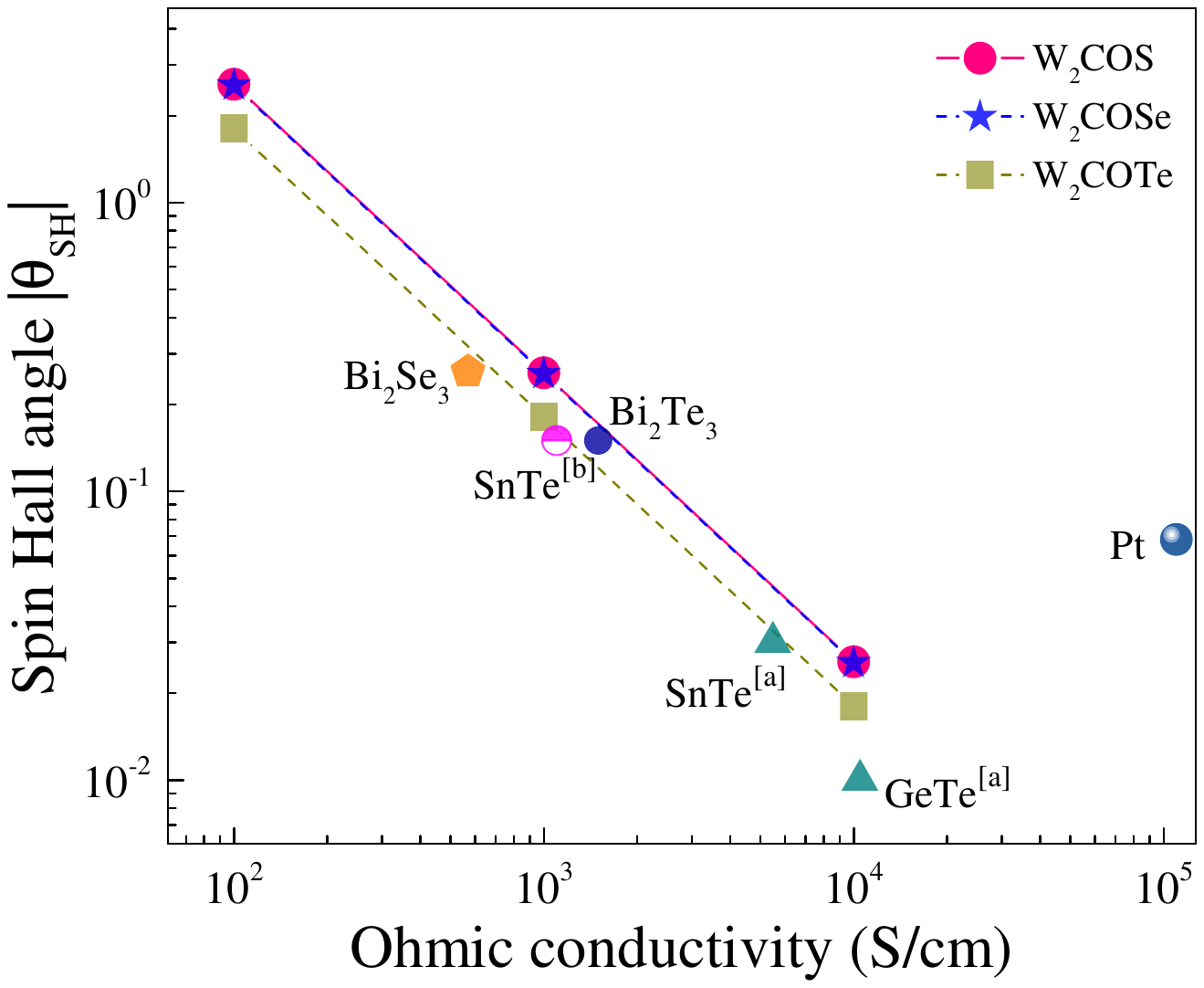}
\caption{Spin Hall angles for W$_2$COS, W$_2$COSe, and W$_2$COTe monolayers computed using ohmic conductivity values ranging from \(10^2\) to \(10^4\) S/cm. The values for other materials, including Bi$_2$Se$_3$~\cite{Farzaneh_PhysRevMaterials.4.114202}, Bi$_2$Te$_3$~\cite{Farzaneh_PhysRevMaterials.4.114202}, GeTe~\cite{Wang_npj_2020}, SnTe~\cite{Wang_npj_2020}, and Pt~\cite{Wang_npj_2020}, are plotted for comparison. Superscript [a] for GeTe and SnTe denotes values corresponding to a carrier concentration of \(2 \times 10^{21}\) $\lvert e \rvert$/cm\(^3\), while superscript [b] represents a carrier concentration of \(3 \times 10^{20}\) $\lvert e \rvert$/cm\(^3\).}
\label{fig: SHA}
\end{figure}

Our calculations reveal that the transverse component \(\sigma_{xy}^z\) of the SHC is 128.4, 127.5, and -90.2 (\(\hbar/e\)) \(\text{S/cm}^{-1}\) for W$_2$COS, W$_2$COSe, and W$_2$COTe monolayers, respectively, at E$_F$. This value remains robust against changes in the broadening parameter within the range of 2-8 meV. These values of SHC at E$_F$ are higher compared to some other 2D MXene, such as V$_2$C ( -25 (\(\hbar/e\)) \(\text{S/cm}^{-1}\)) and Nb$_2$C (-46 (\(\hbar/e\)) \(\text{S/cm}^{-1}\)), and are slightly lower than Ta$_2$C (-187 (\(\hbar/e\)) \(\text{S/cm}^{-1}\)) monolayers~\cite{Zuo_PhysRevB.108.195129}. The longitudinal SHC component \(\sigma_{xx}^x\), on the other hand, is nearly zero at the Fermi level for all three materials. 
Notably, for W\(_2\)COS, \(\sigma_{xy}^z\) spikes to 474.2 (403.9) (\(\hbar/e)\text{S/cm}^{-1}\) at 0.08 eV above the Fermi level, corresponding to a broadening parameter of 2 (4) meV. On the other hand, W\(_2\)COTe shows a \(\sigma_{xy}^z\) value of approximately -338 (\(\hbar/e)\text{S/cm}^{-1}\) below the Fermi level, likely due to the presence of Rashba bands. For W\(_2\)COSe, the SHC peaks at the Fermi level. Despite its low intrinsic SOC and linear Rashba effect, W$_2$COS exhibits the highest SHC value due to its largest cubic Rashba term among the three materials. Analysis of the orbital-projected electronic band structure suggests that the partially occupied W:5d orbitals predominantly contribute to the SHC in these materials.

Although the SHC values in these semiconducting/semimetallic materials are lower than those of heavy metals like platinum or tantalum, their spin Hall angles (SHA) are relatively high and comparable to the theoretically predicted values for some standard topological insulators. This makes them promising candidates for spintronic device design. Due to the lack of experimental data for the exact calculation of spin Hall angles (SHA), we estimated the SHA for all three studied systems using ohmic conductivity (\(\sigma_{yy}\)) values ranging from \(10^2\) to \(10^4\) S/cm, which are typical for semiconducting materials~\cite{Farzaneh_PhysRevMaterials.4.114202,Wang_npj_2020}. 

Applying the formula \(\theta_{SH} = \frac{2e}{\hbar} \left(\lvert\frac{\sigma_{xy}^z}{\sigma_{yy}}\rvert\right)\), we observed that the SHA for W$_2$COTe ranges from 1.8 to 0.018, while for W$_2$COS and W$_2$COSe it varies from 2.5 to 0.025. While these values are comparable to those of other bulk topological insulators such as Bi$_2$Se$_3$ and Bi$_2$Te$_3$~\cite{Farzaneh_PhysRevMaterials.4.114202} (as shown in Figure \ref{fig: SHA}), they are higher than those of narrow bandgap bulk semiconductors GeTe and SnTe~\cite{Wang_npj_2020}, as well as heavy metals such as Pt~\cite{Wang_npj_2020}. Moreover, since the band inversion in these materials is primarily due to crystal field splitting, the SHC can likely be tuned by applying electric fields without altering their topological properties. The relatively high SHA, combined with the tunable SHC, significantly enhances the potential of these materials for practical spintronic applications.

%it is relatively high compared to similar hexagonal 2D materials such as GeTe and SnTe~\cite{Wang_npj_2020}.

%Notably, W\(_2\)COS, as a semiconductor, is expected to exhibit relatively low electrical conductivity, which could lead to a large spin Hall angle advantageous for spintronic device applications. Additionally, since the band inversion in these materials is primarily caused by crystal field splitting, the SHC is likely to be tunable by applying electric fields without altering their topological properties.

\subsection{Conclusions}
In conclusion, we have investigated the electronic and topological properties of a newly predicted class of 2D Rashba materials, W$_2$COX (X = S, Se, Te), derived from the centrosymmetric W$_2$CO$_2$ by substituting an oxygen atom with a chalcogen atom X, thus breaking inversion symmetry. This symmetry breaking introduces Rashba spin splitting, adding a new degree of freedom for manipulating electronic properties of these materials, which enhances the potential of these materials for spintronic applications. Our first-principles DFT calculations confirm that these materials are dynamically and mechanically stable, with mechanical properties comparable to well-known 2D materials like graphene and h-BN.

Without SOC, all three systems are semimetallic. However, SOC opens a band gap, rendering W$_2$COS and W$_2$COSe topologically nontrivial ($Z_2 = 1$) semiconductors, while W$_2$COTe remains topologically trivial ($Z_2 = 0$). The broken inversion symmetry leads to pronounced nonlinear Rashba spin splitting in all three systems. 
%Due to the C$_{3v}$ point group symmetry and deviations from a purely parabolic band structure, the linear Rashba-Bychkov model alone does not accurately capture the observed Rashba splitting, necessitating the inclusion of $k^3$ terms in the Rashba Hamiltonian. Our calculations reveal that these materials exhibit significant nonlinear RSOC. 
Notably, despite having the lowest linear Rashba constant, W$_2$COS exhibits the highest $k^3$ contribution of -45.9 eV\AA$^3$ near the Fermi level (E$_F$). This cubic term also causes a warping effect in the spin textures of these materials.

Furthermore, the studied monolayers exhibit substantial SHC, with W$_2$COS achieving the highest SHC of approximately 474 (\(\hbar/e) \text{S/cm}^{-1}\) around 80 meV above the Fermi level owing to dominant $k^3$ contributions. While the SHC of W$_2$COX materials is lower than that of heavy metals like platinum and tantalum, these materials exhibit relatively large spin Hall angles (0.018-2.5 at E$_F$), comparable to other bulk topological insulators and exceeding some narrow bandgap semiconductors. The combination of high SHC, large spin Hall angles, and the ability to tune SHC via electric fields without altering the inherent topological properties, which arise from crystal field splitting, highlights the potential of W$_2$COX materials for advanced spintronic applications.

\section*{Acknowledgements}
Authors acknowledge support from the University Research Awards at the University of Rochester. SS is supported by the U.S.~Department of Energy, Office of Science, Office of Fusion Energy Sciences, Quantum Information Science program under Award No.~DE-SC-0020340. 
Authors thank the Pittsburgh Supercomputer Center (Bridges2) supported by the Advanced Cyberinfrastructure Coordination Ecosystem: Services \& Support (ACCESS) program, which is supported by National Science Foundation grants \#2138259, \#2138286, \#2138307, \#2137603, and \#2138296. 

\medskip

\appendix
%\label{Appendix A}

\section{Crystallographic parameters and mechanical constants}
\label{Appendix A}

Appendix \ref{Appendix A} includes the list of crystallographic parameters and mechanical constants of W$_2$COX (X = S, Se, Te) monolayers.
\begin{table}[h]
\caption{List of crystallographic parameters and mechanical constants of W$_2$COX (X = S, Se, Te) monolayers} \label{tab:crystal parameters} 
\begin{tabular}{cccc}
\hline
Parameters & W$_2$COS & W$_2$COSe & W$_2$COTe \\
\hline
\textit{a = b} (\AA) & 2.94 & 2.96 & 3.03 \\
W-X (\AA) & 2.4 & 2.5 & 2.7 \\
W-W (\AA) & 2.94 & 2.96 & 3.0 \\
W-C (\AA) & 2.2 & 2.2 & 2.2 \\
W-O (\AA) & 2.1 & 2.1 & 2.1 \\
h (\AA) & 1.6 & 1.8 & 2.1 \\
C$_{11}$ (N/m) & 381.8 & 374.9 & 327.4 \\
C$_{12}$ (N/m) & 121.4 & 104.3 & 81.1 \\
Young's modulus (N/m) & 343.4 & 345.9 & 307.9 \\
Shear modulus (N/m) & 252.3 & 239.8 & 204.4 \\
2D layer modulus (N/m) & 131.4 & 136.6 & 123.1 \\
\hline
\end{tabular}
\end{table}

\begin{figure}[h]
\centering
\includegraphics[trim=0.2cm 0cm 2.5cm 0.2cm, clip=true, width=1\columnwidth]{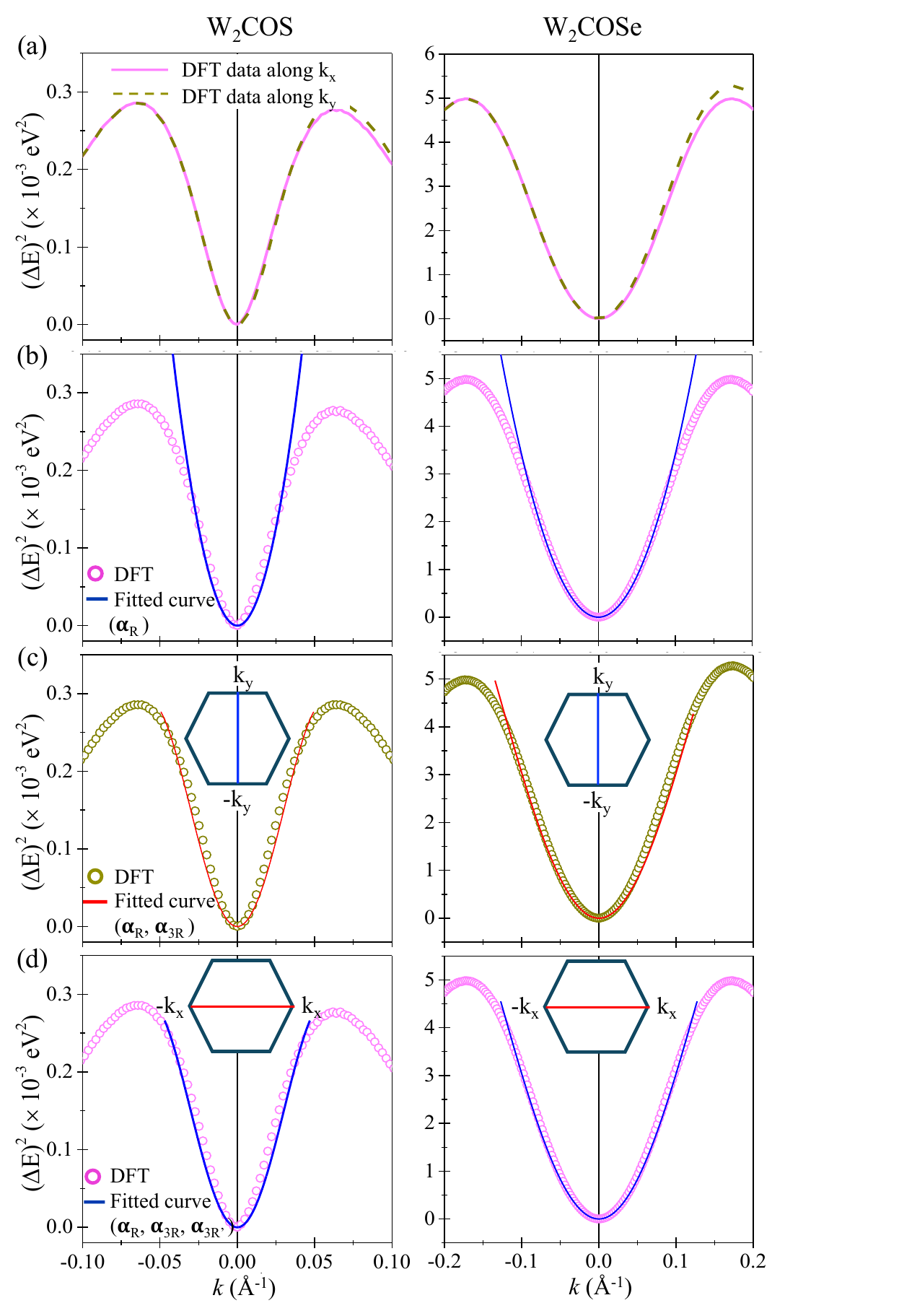}
\caption{(a) Square of the energy splitting $\Delta E = [E_+(k) - E_-(k)] / 2$  of Rashba bands as a function of \(k\). The magenta line represents data along the $k_x$ direction, while the dark yellow dashed line represents data along the $k_y$  direction. (b)-(d) Fitting of DFT data to various functions to extract first-order and third-order Rashba parameters. Solid lines represent the fitted curves, and circular markers indicate the DFT data. The left panel shows results for W$_2$COS, and the right panel shows results for W$_2$COSe monolayers. }
\label{fig: W2COS/Se_higher_order}
\end{figure}
\section{Nonlinear RSOC in W$_2$COS and W$_2$COSe }
\label{Appendix B}

Appendix \ref{Appendix B} presents the fitting of DFT data to various functions to determine the linear and third-order Rashba parameters for W$_2$COS and W$_2$COSe monolayers (Figure \ref{fig: W2COS/Se_higher_order}).

\bibliography{bibfile}% Produces the bibliography via BibTeX.

\begin{thebibliography}{62}%
\makeatletter
\providecommand \@ifxundefined [1]{%
 \@ifx{#1\undefined}
}%
\providecommand \@ifnum [1]{%
 \ifnum #1\expandafter \@firstoftwo
 \else \expandafter \@secondoftwo
 \fi
}%
\providecommand \@ifx [1]{%
 \ifx #1\expandafter \@firstoftwo
 \else \expandafter \@secondoftwo
 \fi
}%
\providecommand \natexlab [1]{#1}%
\providecommand \enquote  [1]{``#1''}%
\providecommand \bibnamefont  [1]{#1}%
\providecommand \bibfnamefont [1]{#1}%
\providecommand \citenamefont [1]{#1}%
\providecommand \href@noop [0]{\@secondoftwo}%
\providecommand \href [0]{\begingroup \@sanitize@url \@href}%
\providecommand \@href[1]{\@@startlink{#1}\@@href}%
\providecommand \@@href[1]{\endgroup#1\@@endlink}%
\providecommand \@sanitize@url [0]{\catcode `\\12\catcode `\$12\catcode
  `\&12\catcode `\#12\catcode `\^12\catcode `\_12\catcode `\%12\relax}%
\providecommand \@@startlink[1]{}%
\providecommand \@@endlink[0]{}%
\providecommand \url  [0]{\begingroup\@sanitize@url \@url }%
\providecommand \@url [1]{\endgroup\@href {#1}{\urlprefix }}%
\providecommand \urlprefix  [0]{URL }%
\providecommand \Eprint [0]{\href }%
\providecommand \doibase [0]{http://dx.doi.org/}%
\providecommand \selectlanguage [0]{\@gobble}%
\providecommand \bibinfo  [0]{\@secondoftwo}%
\providecommand \bibfield  [0]{\@secondoftwo}%
\providecommand \translation [1]{[#1]}%
\providecommand \BibitemOpen [0]{}%
\providecommand \bibitemStop [0]{}%
\providecommand \bibitemNoStop [0]{.\EOS\space}%
\providecommand \EOS [0]{\spacefactor3000\relax}%
\providecommand \BibitemShut  [1]{\csname bibitem#1\endcsname}%
\let\auto@bib@innerbib\@empty
%</preamble>
\bibitem [{\citenamefont {Murakami}\ \emph {et~al.}(2003)\citenamefont
  {Murakami}, \citenamefont {Nagaosa},\ and\ \citenamefont
  {Zhang}}]{Murakami_Science_2003}%
  \BibitemOpen
  \bibfield  {author} {\bibinfo {author} {\bibfnamefont {S.}~\bibnamefont
  {Murakami}}, \bibinfo {author} {\bibfnamefont {N.}~\bibnamefont {Nagaosa}}, \
  and\ \bibinfo {author} {\bibfnamefont {S.-C.}\ \bibnamefont {Zhang}},\ }\href
  {\doibase 10.1126/science.1087128} {\bibfield  {journal} {\bibinfo  {journal}
  {Science}\ }\textbf {\bibinfo {volume} {301}},\ \bibinfo {pages} {1348}
  (\bibinfo {year} {2003})}\BibitemShut {NoStop}%
\bibitem [{\citenamefont {Qi}\ and\ \citenamefont
  {Zhang}(2011)}]{qi2011_rev_mod_phy}%
  \BibitemOpen
  \bibfield  {author} {\bibinfo {author} {\bibfnamefont {X.-L.}\ \bibnamefont
  {Qi}}\ and\ \bibinfo {author} {\bibfnamefont {S.-C.}\ \bibnamefont {Zhang}},\
  }\href@noop {} {\bibfield  {journal} {\bibinfo  {journal} {Reviews of Modern
  Physics}\ }\textbf {\bibinfo {volume} {83}},\ \bibinfo {pages} {1057}
  (\bibinfo {year} {2011})}\BibitemShut {NoStop}%
\bibitem [{\citenamefont {Tang}\ \emph {et~al.}(2017)\citenamefont {Tang},
  \citenamefont {Zhang}, \citenamefont {Wong}, \citenamefont {Pedramrazi},
  \citenamefont {Tsai}, \citenamefont {Jia}, \citenamefont {Moritz},
  \citenamefont {Claassen}, \citenamefont {Ryu}, \citenamefont {Kahn},
  \citenamefont {Jiang}, \citenamefont {Yan}, \citenamefont {Hashimoto},
  \citenamefont {Lu}, \citenamefont {Moore}, \citenamefont {Hwang},
  \citenamefont {Hwang}, \citenamefont {Hussain}, \citenamefont {Chen},
  \citenamefont {Ugeda}, \citenamefont {Liu}, \citenamefont {Xie},
  \citenamefont {Devereaux}, \citenamefont {Crommie}, \citenamefont {Mo},\ and\
  \citenamefont {Shen}}]{Tang2017_nature}%
  \BibitemOpen
  \bibfield  {author} {\bibinfo {author} {\bibfnamefont {S.}~\bibnamefont
  {Tang}}, \bibinfo {author} {\bibfnamefont {C.}~\bibnamefont {Zhang}},
  \bibinfo {author} {\bibfnamefont {D.}~\bibnamefont {Wong}}, \bibinfo {author}
  {\bibfnamefont {Z.}~\bibnamefont {Pedramrazi}}, \bibinfo {author}
  {\bibfnamefont {H.-Z.}\ \bibnamefont {Tsai}}, \bibinfo {author}
  {\bibfnamefont {C.}~\bibnamefont {Jia}}, \bibinfo {author} {\bibfnamefont
  {B.}~\bibnamefont {Moritz}}, \bibinfo {author} {\bibfnamefont
  {M.}~\bibnamefont {Claassen}}, \bibinfo {author} {\bibfnamefont
  {H.}~\bibnamefont {Ryu}}, \bibinfo {author} {\bibfnamefont {S.}~\bibnamefont
  {Kahn}}, \bibinfo {author} {\bibfnamefont {J.}~\bibnamefont {Jiang}},
  \bibinfo {author} {\bibfnamefont {H.}~\bibnamefont {Yan}}, \bibinfo {author}
  {\bibfnamefont {M.}~\bibnamefont {Hashimoto}}, \bibinfo {author}
  {\bibfnamefont {D.}~\bibnamefont {Lu}}, \bibinfo {author} {\bibfnamefont
  {R.~G.}\ \bibnamefont {Moore}}, \bibinfo {author} {\bibfnamefont {C.-C.}\
  \bibnamefont {Hwang}}, \bibinfo {author} {\bibfnamefont {C.}~\bibnamefont
  {Hwang}}, \bibinfo {author} {\bibfnamefont {Z.}~\bibnamefont {Hussain}},
  \bibinfo {author} {\bibfnamefont {Y.}~\bibnamefont {Chen}}, \bibinfo {author}
  {\bibfnamefont {M.~M.}\ \bibnamefont {Ugeda}}, \bibinfo {author}
  {\bibfnamefont {Z.}~\bibnamefont {Liu}}, \bibinfo {author} {\bibfnamefont
  {X.}~\bibnamefont {Xie}}, \bibinfo {author} {\bibfnamefont {T.~P.}\
  \bibnamefont {Devereaux}}, \bibinfo {author} {\bibfnamefont {M.~F.}\
  \bibnamefont {Crommie}}, \bibinfo {author} {\bibfnamefont {S.-K.}\
  \bibnamefont {Mo}}, \ and\ \bibinfo {author} {\bibfnamefont {Z.-X.}\
  \bibnamefont {Shen}},\ }\href {\doibase 10.1038/nphys4174} {\bibfield
  {journal} {\bibinfo  {journal} {Nature Physics}\ }\textbf {\bibinfo {volume}
  {13}},\ \bibinfo {pages} {683} (\bibinfo {year} {2017})}\BibitemShut
  {NoStop}%
\bibitem [{\citenamefont {Sau}\ \emph {et~al.}(2010{\natexlab{a}})\citenamefont
  {Sau}, \citenamefont {Lutchyn}, \citenamefont {Tewari},\ and\ \citenamefont
  {Das~Sarma}}]{Sau_PhysRevLett.104.040502}%
  \BibitemOpen
  \bibfield  {author} {\bibinfo {author} {\bibfnamefont {J.~D.}\ \bibnamefont
  {Sau}}, \bibinfo {author} {\bibfnamefont {R.~M.}\ \bibnamefont {Lutchyn}},
  \bibinfo {author} {\bibfnamefont {S.}~\bibnamefont {Tewari}}, \ and\ \bibinfo
  {author} {\bibfnamefont {S.}~\bibnamefont {Das~Sarma}},\ }\href {\doibase
  10.1103/PhysRevLett.104.040502} {\bibfield  {journal} {\bibinfo  {journal}
  {Phys. Rev. Lett.}\ }\textbf {\bibinfo {volume} {104}},\ \bibinfo {pages}
  {040502} (\bibinfo {year} {2010}{\natexlab{a}})}\BibitemShut {NoStop}%
\bibitem [{\citenamefont {Sau}\ \emph {et~al.}(2010{\natexlab{b}})\citenamefont
  {Sau}, \citenamefont {Tewari}, \citenamefont {Lutchyn}, \citenamefont
  {Stanescu},\ and\ \citenamefont {Das~Sarma}}]{Sau_PhysRevB.82.214509}%
  \BibitemOpen
  \bibfield  {author} {\bibinfo {author} {\bibfnamefont {J.~D.}\ \bibnamefont
  {Sau}}, \bibinfo {author} {\bibfnamefont {S.}~\bibnamefont {Tewari}},
  \bibinfo {author} {\bibfnamefont {R.~M.}\ \bibnamefont {Lutchyn}}, \bibinfo
  {author} {\bibfnamefont {T.~D.}\ \bibnamefont {Stanescu}}, \ and\ \bibinfo
  {author} {\bibfnamefont {S.}~\bibnamefont {Das~Sarma}},\ }\href {\doibase
  10.1103/PhysRevB.82.214509} {\bibfield  {journal} {\bibinfo  {journal} {Phys.
  Rev. B}\ }\textbf {\bibinfo {volume} {82}},\ \bibinfo {pages} {214509}
  (\bibinfo {year} {2010}{\natexlab{b}})}\BibitemShut {NoStop}%
\bibitem [{\citenamefont {Weng}\ \emph {et~al.}(2014)\citenamefont {Weng},
  \citenamefont {Dai},\ and\ \citenamefont {Fang}}]{Weng_MRS_Bulletin_2014}%
  \BibitemOpen
  \bibfield  {author} {\bibinfo {author} {\bibfnamefont {H.}~\bibnamefont
  {Weng}}, \bibinfo {author} {\bibfnamefont {X.}~\bibnamefont {Dai}}, \ and\
  \bibinfo {author} {\bibfnamefont {Z.}~\bibnamefont {Fang}},\ }\href {\doibase
  10.1557/mrs.2014.216} {\bibfield  {journal} {\bibinfo  {journal} {MRS
  Bulletin}\ }\textbf {\bibinfo {volume} {39}},\ \bibinfo {pages} {849}
  (\bibinfo {year} {2014})}\BibitemShut {NoStop}%
\bibitem [{\citenamefont {Ando}(2013)}]{Ando_Journal_of_phy_soc_Japan}%
  \BibitemOpen
  \bibfield  {author} {\bibinfo {author} {\bibfnamefont {Y.}~\bibnamefont
  {Ando}},\ }\href {\doibase 10.7566/JPSJ.82.102001} {\bibfield  {journal}
  {\bibinfo  {journal} {Journal of the Physical Society of Japan}\ }\textbf
  {\bibinfo {volume} {82}},\ \bibinfo {pages} {102001} (\bibinfo {year}
  {2013})}\BibitemShut {NoStop}%
\bibitem [{\citenamefont {Bernevig}\ \emph {et~al.}(2006)\citenamefont
  {Bernevig}, \citenamefont {Hughes},\ and\ \citenamefont
  {Zhang}}]{bernevig2006_science}%
  \BibitemOpen
  \bibfield  {author} {\bibinfo {author} {\bibfnamefont {B.~A.}\ \bibnamefont
  {Bernevig}}, \bibinfo {author} {\bibfnamefont {T.~L.}\ \bibnamefont
  {Hughes}}, \ and\ \bibinfo {author} {\bibfnamefont {S.-C.}\ \bibnamefont
  {Zhang}},\ }\href@noop {} {\bibfield  {journal} {\bibinfo  {journal}
  {science}\ }\textbf {\bibinfo {volume} {314}},\ \bibinfo {pages} {1757}
  (\bibinfo {year} {2006})}\BibitemShut {NoStop}%
\bibitem [{\citenamefont {K{\"o}nig}\ \emph {et~al.}(2007)\citenamefont
  {K{\"o}nig}, \citenamefont {Wiedmann}, \citenamefont {Br{\"u}ne},
  \citenamefont {Roth}, \citenamefont {Buhmann},\ and\ \citenamefont
  {Molenkamp}}]{konig2007_science}%
  \BibitemOpen
  \bibfield  {author} {\bibinfo {author} {\bibfnamefont {M.}~\bibnamefont
  {K{\"o}nig}}, \bibinfo {author} {\bibfnamefont {S.}~\bibnamefont {Wiedmann}},
  \bibinfo {author} {\bibfnamefont {C.}~\bibnamefont {Br{\"u}ne}}, \bibinfo
  {author} {\bibfnamefont {A.}~\bibnamefont {Roth}}, \bibinfo {author}
  {\bibfnamefont {H.}~\bibnamefont {Buhmann}}, \ and\ \bibinfo {author}
  {\bibfnamefont {L.}~\bibnamefont {Molenkamp}},\ }\href@noop {} {\bibfield
  {journal} {\bibinfo  {journal} {Science}\ }\textbf {\bibinfo {volume}
  {318}},\ \bibinfo {pages} {766} (\bibinfo {year} {2007})}\BibitemShut
  {NoStop}%
\bibitem [{\citenamefont {K{\"o}nig}\ \emph {et~al.}(2008)\citenamefont
  {K{\"o}nig}, \citenamefont {Buhmann}, \citenamefont {W.~Molenkamp},
  \citenamefont {Hughes}, \citenamefont {Liu}, \citenamefont {Qi},\ and\
  \citenamefont {Zhang}}]{konig2008_JPPJ}%
  \BibitemOpen
  \bibfield  {author} {\bibinfo {author} {\bibfnamefont {M.}~\bibnamefont
  {K{\"o}nig}}, \bibinfo {author} {\bibfnamefont {H.}~\bibnamefont {Buhmann}},
  \bibinfo {author} {\bibfnamefont {L.}~\bibnamefont {W.~Molenkamp}}, \bibinfo
  {author} {\bibfnamefont {T.}~\bibnamefont {Hughes}}, \bibinfo {author}
  {\bibfnamefont {C.-X.}\ \bibnamefont {Liu}}, \bibinfo {author} {\bibfnamefont
  {X.-L.}\ \bibnamefont {Qi}}, \ and\ \bibinfo {author} {\bibfnamefont {S.-C.}\
  \bibnamefont {Zhang}},\ }\href@noop {} {\bibfield  {journal} {\bibinfo
  {journal} {Journal of the Physical Society of Japan}\ }\textbf {\bibinfo
  {volume} {77}},\ \bibinfo {pages} {031007} (\bibinfo {year}
  {2008})}\BibitemShut {NoStop}%
\bibitem [{\citenamefont {Ren}\ \emph {et~al.}(2016)\citenamefont {Ren},
  \citenamefont {Qiao},\ and\ \citenamefont
  {Niu}}]{ren2016_REPRTS_ON_PROGRESS_PHYSICS}%
  \BibitemOpen
  \bibfield  {author} {\bibinfo {author} {\bibfnamefont {Y.}~\bibnamefont
  {Ren}}, \bibinfo {author} {\bibfnamefont {Z.}~\bibnamefont {Qiao}}, \ and\
  \bibinfo {author} {\bibfnamefont {Q.}~\bibnamefont {Niu}},\ }\href@noop {}
  {\bibfield  {journal} {\bibinfo  {journal} {Reports on Progress in Physics}\
  }\textbf {\bibinfo {volume} {79}},\ \bibinfo {pages} {066501} (\bibinfo
  {year} {2016})}\BibitemShut {NoStop}%
\bibitem [{\citenamefont {Singh}\ \emph {et~al.}(2019)\citenamefont {Singh},
  \citenamefont {Zanolli}, \citenamefont {Amsler}, \citenamefont {Belhadji},
  \citenamefont {Sofo}, \citenamefont {Verstraete},\ and\ \citenamefont
  {Romero}}]{SinghJPCL2019}%
  \BibitemOpen
  \bibfield  {author} {\bibinfo {author} {\bibfnamefont {S.}~\bibnamefont
  {Singh}}, \bibinfo {author} {\bibfnamefont {Z.}~\bibnamefont {Zanolli}},
  \bibinfo {author} {\bibfnamefont {M.}~\bibnamefont {Amsler}}, \bibinfo
  {author} {\bibfnamefont {B.}~\bibnamefont {Belhadji}}, \bibinfo {author}
  {\bibfnamefont {J.~O.}\ \bibnamefont {Sofo}}, \bibinfo {author}
  {\bibfnamefont {M.~J.}\ \bibnamefont {Verstraete}}, \ and\ \bibinfo {author}
  {\bibfnamefont {A.~H.}\ \bibnamefont {Romero}},\ }\href {\doibase
  10.1021/acs.jpclett.9b03043} {\bibfield  {journal} {\bibinfo  {journal} {The
  Journal of Physical Chemistry Letters}\ }\textbf {\bibinfo {volume} {10}},\
  \bibinfo {pages} {7324} (\bibinfo {year} {2019})}\BibitemShut {NoStop}%
\bibitem [{\citenamefont {Roche}\ \emph {et~al.}(2024)\citenamefont {Roche},
  \citenamefont {van Wees}, \citenamefont {Garello},\ and\ \citenamefont
  {Valenzuela}}]{Roche_2024}%
  \BibitemOpen
  \bibfield  {author} {\bibinfo {author} {\bibfnamefont {S.}~\bibnamefont
  {Roche}}, \bibinfo {author} {\bibfnamefont {B.}~\bibnamefont {van Wees}},
  \bibinfo {author} {\bibfnamefont {K.}~\bibnamefont {Garello}}, \ and\
  \bibinfo {author} {\bibfnamefont {S.~O.}\ \bibnamefont {Valenzuela}},\ }\href
  {\doibase 10.1088/2053-1583/ad64e2} {\bibfield  {journal} {\bibinfo
  {journal} {2D Materials}\ }\textbf {\bibinfo {volume} {11}},\ \bibinfo
  {pages} {043001} (\bibinfo {year} {2024})}\BibitemShut {NoStop}%
\bibitem [{\citenamefont {Weng}\ \emph {et~al.}(2015)\citenamefont {Weng},
  \citenamefont {Ranjbar}, \citenamefont {Liang}, \citenamefont {Song},
  \citenamefont {Khazaei}, \citenamefont {Yunoki}, \citenamefont {Arai},
  \citenamefont {Kawazoe}, \citenamefont {Fang},\ and\ \citenamefont
  {Dai}}]{Weng_PhysRevB.92.075436}%
  \BibitemOpen
  \bibfield  {author} {\bibinfo {author} {\bibfnamefont {H.}~\bibnamefont
  {Weng}}, \bibinfo {author} {\bibfnamefont {A.}~\bibnamefont {Ranjbar}},
  \bibinfo {author} {\bibfnamefont {Y.}~\bibnamefont {Liang}}, \bibinfo
  {author} {\bibfnamefont {Z.}~\bibnamefont {Song}}, \bibinfo {author}
  {\bibfnamefont {M.}~\bibnamefont {Khazaei}}, \bibinfo {author} {\bibfnamefont
  {S.}~\bibnamefont {Yunoki}}, \bibinfo {author} {\bibfnamefont
  {M.}~\bibnamefont {Arai}}, \bibinfo {author} {\bibfnamefont {Y.}~\bibnamefont
  {Kawazoe}}, \bibinfo {author} {\bibfnamefont {Z.}~\bibnamefont {Fang}}, \
  and\ \bibinfo {author} {\bibfnamefont {X.}~\bibnamefont {Dai}},\ }\href
  {\doibase 10.1103/PhysRevB.92.075436} {\bibfield  {journal} {\bibinfo
  {journal} {Phys. Rev. B}\ }\textbf {\bibinfo {volume} {92}},\ \bibinfo
  {pages} {075436} (\bibinfo {year} {2015})}\BibitemShut {NoStop}%
\bibitem [{\citenamefont
  {Barsoum}(2000)}]{Barsoum_progree_in_solid_state_chem}%
  \BibitemOpen
  \bibfield  {author} {\bibinfo {author} {\bibfnamefont {M.~W.}\ \bibnamefont
  {Barsoum}},\ }\href {\doibase https://doi.org/10.1016/S0079-6786(00)00006-6}
  {\bibfield  {journal} {\bibinfo  {journal} {Progress in Solid State
  Chemistry}\ }\textbf {\bibinfo {volume} {28}},\ \bibinfo {pages} {201}
  (\bibinfo {year} {2000})}\BibitemShut {NoStop}%
\bibitem [{\citenamefont {Harris}\ \emph {et~al.}(2015)\citenamefont {Harris},
  \citenamefont {Bugnet}, \citenamefont {Naguib}, \citenamefont {Barsoum},\
  and\ \citenamefont {Goward}}]{Harris_Jornal-of_phy_chem_c_2015}%
  \BibitemOpen
  \bibfield  {author} {\bibinfo {author} {\bibfnamefont {K.~J.}\ \bibnamefont
  {Harris}}, \bibinfo {author} {\bibfnamefont {M.}~\bibnamefont {Bugnet}},
  \bibinfo {author} {\bibfnamefont {M.}~\bibnamefont {Naguib}}, \bibinfo
  {author} {\bibfnamefont {M.~W.}\ \bibnamefont {Barsoum}}, \ and\ \bibinfo
  {author} {\bibfnamefont {G.~R.}\ \bibnamefont {Goward}},\ }\href {\doibase
  10.1021/acs.jpcc.5b03038} {\bibfield  {journal} {\bibinfo  {journal} {The
  Journal of Physical Chemistry C}\ }\textbf {\bibinfo {volume} {119}},\
  \bibinfo {pages} {13713} (\bibinfo {year} {2015})}\BibitemShut {NoStop}%
\bibitem [{\citenamefont {Khazaei}\ \emph {et~al.}(2013)\citenamefont
  {Khazaei}, \citenamefont {Arai}, \citenamefont {Sasaki}, \citenamefont
  {Chung}, \citenamefont {Venkataramanan}, \citenamefont {Estili},
  \citenamefont {Sakka},\ and\ \citenamefont
  {Kawazoe}}]{Khazaei_Adv_functional_mat}%
  \BibitemOpen
  \bibfield  {author} {\bibinfo {author} {\bibfnamefont {M.}~\bibnamefont
  {Khazaei}}, \bibinfo {author} {\bibfnamefont {M.}~\bibnamefont {Arai}},
  \bibinfo {author} {\bibfnamefont {T.}~\bibnamefont {Sasaki}}, \bibinfo
  {author} {\bibfnamefont {C.-Y.}\ \bibnamefont {Chung}}, \bibinfo {author}
  {\bibfnamefont {N.~S.}\ \bibnamefont {Venkataramanan}}, \bibinfo {author}
  {\bibfnamefont {M.}~\bibnamefont {Estili}}, \bibinfo {author} {\bibfnamefont
  {Y.}~\bibnamefont {Sakka}}, \ and\ \bibinfo {author} {\bibfnamefont
  {Y.}~\bibnamefont {Kawazoe}},\ }\href {\doibase
  https://doi.org/10.1002/adfm.201202502} {\bibfield  {journal} {\bibinfo
  {journal} {Advanced Functional Materials}\ }\textbf {\bibinfo {volume}
  {23}},\ \bibinfo {pages} {2185} (\bibinfo {year} {2013})}\BibitemShut
  {NoStop}%
\bibitem [{\citenamefont {Khazaei}\ \emph {et~al.}(2014)\citenamefont
  {Khazaei}, \citenamefont {Arai}, \citenamefont {Sasaki}, \citenamefont
  {Estili},\ and\ \citenamefont {Sakka}}]{Khazaei_phys_chem_chem_phys_2014}%
  \BibitemOpen
  \bibfield  {author} {\bibinfo {author} {\bibfnamefont {M.}~\bibnamefont
  {Khazaei}}, \bibinfo {author} {\bibfnamefont {M.}~\bibnamefont {Arai}},
  \bibinfo {author} {\bibfnamefont {T.}~\bibnamefont {Sasaki}}, \bibinfo
  {author} {\bibfnamefont {M.}~\bibnamefont {Estili}}, \ and\ \bibinfo {author}
  {\bibfnamefont {Y.}~\bibnamefont {Sakka}},\ }\href {\doibase
  10.1039/C4CP00467A} {\bibfield  {journal} {\bibinfo  {journal} {Phys. Chem.
  Chem. Phys.}\ }\textbf {\bibinfo {volume} {16}},\ \bibinfo {pages} {7841}
  (\bibinfo {year} {2014})}\BibitemShut {NoStop}%
\bibitem [{\citenamefont {Si}\ \emph {et~al.}(2016)\citenamefont {Si},
  \citenamefont {Jin}, \citenamefont {Zhou}, \citenamefont {Sun},\ and\
  \citenamefont {Liu}}]{Si_Nano_Letters_2016}%
  \BibitemOpen
  \bibfield  {author} {\bibinfo {author} {\bibfnamefont {C.}~\bibnamefont
  {Si}}, \bibinfo {author} {\bibfnamefont {K.-H.}\ \bibnamefont {Jin}},
  \bibinfo {author} {\bibfnamefont {J.}~\bibnamefont {Zhou}}, \bibinfo {author}
  {\bibfnamefont {Z.}~\bibnamefont {Sun}}, \ and\ \bibinfo {author}
  {\bibfnamefont {F.}~\bibnamefont {Liu}},\ }\href {\doibase
  10.1021/acs.nanolett.6b03118} {\bibfield  {journal} {\bibinfo  {journal}
  {Nano Letters}\ }\textbf {\bibinfo {volume} {16}},\ \bibinfo {pages} {6584}
  (\bibinfo {year} {2016})}\BibitemShut {NoStop}%
\bibitem [{\citenamefont {Yang}\ \emph {et~al.}(2022)\citenamefont {Yang},
  \citenamefont {Wang}, \citenamefont {Liu}, \citenamefont {Fang},
  \citenamefont {Chen},\ and\ \citenamefont
  {Cheng}}]{Yang_J_Mater_Chem_A_D2TA07161D}%
  \BibitemOpen
  \bibfield  {author} {\bibinfo {author} {\bibfnamefont {T.}~\bibnamefont
  {Yang}}, \bibinfo {author} {\bibfnamefont {Q.}~\bibnamefont {Wang}}, \bibinfo
  {author} {\bibfnamefont {Z.}~\bibnamefont {Liu}}, \bibinfo {author}
  {\bibfnamefont {J.}~\bibnamefont {Fang}}, \bibinfo {author} {\bibfnamefont
  {X.}~\bibnamefont {Chen}}, \ and\ \bibinfo {author} {\bibfnamefont
  {X.}~\bibnamefont {Cheng}},\ }\href {\doibase 10.1039/D2TA07161D} {\bibfield
  {journal} {\bibinfo  {journal} {J. Mater. Chem. A}\ }\textbf {\bibinfo
  {volume} {10}},\ \bibinfo {pages} {24238} (\bibinfo {year}
  {2022})}\BibitemShut {NoStop}%
\bibitem [{\citenamefont {Yang}\ \emph {et~al.}(2024)\citenamefont {Yang},
  \citenamefont {Liu}, \citenamefont {Fang}, \citenamefont {Liu}, \citenamefont
  {Qiao}, \citenamefont {Zhu}, \citenamefont {Cheng}, \citenamefont {Zhang},\
  and\ \citenamefont {Chen}}]{Yang_Phy_Chem_Chem_Phy_D3CP05142K}%
  \BibitemOpen
  \bibfield  {author} {\bibinfo {author} {\bibfnamefont {T.}~\bibnamefont
  {Yang}}, \bibinfo {author} {\bibfnamefont {X.}~\bibnamefont {Liu}}, \bibinfo
  {author} {\bibfnamefont {J.}~\bibnamefont {Fang}}, \bibinfo {author}
  {\bibfnamefont {Z.}~\bibnamefont {Liu}}, \bibinfo {author} {\bibfnamefont
  {Z.}~\bibnamefont {Qiao}}, \bibinfo {author} {\bibfnamefont {Z.}~\bibnamefont
  {Zhu}}, \bibinfo {author} {\bibfnamefont {Q.}~\bibnamefont {Cheng}}, \bibinfo
  {author} {\bibfnamefont {Y.}~\bibnamefont {Zhang}}, \ and\ \bibinfo {author}
  {\bibfnamefont {X.}~\bibnamefont {Chen}},\ }\href {\doibase
  10.1039/D3CP05142K} {\bibfield  {journal} {\bibinfo  {journal} {Phys. Chem.
  Chem. Phys.}\ }\textbf {\bibinfo {volume} {26}},\ \bibinfo {pages} {7475}
  (\bibinfo {year} {2024})}\BibitemShut {NoStop}%
\bibitem [{\citenamefont {Maghirang}\ \emph {et~al.}(2022)\citenamefont
  {Maghirang}, \citenamefont {Macam}, \citenamefont {Sufyan}, \citenamefont
  {Huang}, \citenamefont {Hsu},\ and\ \citenamefont
  {Chuang}}]{MAGHIRANG_chinese_journal_of_phy_20222346}%
  \BibitemOpen
  \bibfield  {author} {\bibinfo {author} {\bibfnamefont {A.~B.}\ \bibnamefont
  {Maghirang}}, \bibinfo {author} {\bibfnamefont {G.}~\bibnamefont {Macam}},
  \bibinfo {author} {\bibfnamefont {A.}~\bibnamefont {Sufyan}}, \bibinfo
  {author} {\bibfnamefont {Z.-Q.}\ \bibnamefont {Huang}}, \bibinfo {author}
  {\bibfnamefont {C.-H.}\ \bibnamefont {Hsu}}, \ and\ \bibinfo {author}
  {\bibfnamefont {F.-C.}\ \bibnamefont {Chuang}},\ }\href {\doibase
  https://doi.org/10.1016/j.cjph.2022.04.012} {\bibfield  {journal} {\bibinfo
  {journal} {Chinese Journal of Physics}\ }\textbf {\bibinfo {volume} {77}},\
  \bibinfo {pages} {2346} (\bibinfo {year} {2022})}\BibitemShut {NoStop}%
\bibitem [{\citenamefont {Heide}\ \emph {et~al.}(2006)\citenamefont {Heide},
  \citenamefont {Bihlmayer}, \citenamefont {Mavropoulos}, \citenamefont
  {Bringer},\ and\ \citenamefont {Bl{\"u}gel}}]{heide2006spin}%
  \BibitemOpen
  \bibfield  {author} {\bibinfo {author} {\bibfnamefont {M.}~\bibnamefont
  {Heide}}, \bibinfo {author} {\bibfnamefont {G.}~\bibnamefont {Bihlmayer}},
  \bibinfo {author} {\bibfnamefont {P.}~\bibnamefont {Mavropoulos}}, \bibinfo
  {author} {\bibfnamefont {A.}~\bibnamefont {Bringer}}, \ and\ \bibinfo
  {author} {\bibfnamefont {S.}~\bibnamefont {Bl{\"u}gel}},\ }\href@noop {}
  {\bibfield  {journal} {\bibinfo  {journal} {Newsletter of the Psi-K Network}\
  }\textbf {\bibinfo {volume} {78}} (\bibinfo {year} {2006})}\BibitemShut
  {NoStop}%
\bibitem [{\citenamefont {Bihlmayer}\ \emph {et~al.}(2015)\citenamefont
  {Bihlmayer}, \citenamefont {Rader},\ and\ \citenamefont
  {Winkler}}]{Bihlmayer_2015}%
  \BibitemOpen
  \bibfield  {author} {\bibinfo {author} {\bibfnamefont {G.}~\bibnamefont
  {Bihlmayer}}, \bibinfo {author} {\bibfnamefont {O.}~\bibnamefont {Rader}}, \
  and\ \bibinfo {author} {\bibfnamefont {R.}~\bibnamefont {Winkler}},\ }\href
  {\doibase 10.1088/1367-2630/17/5/050202} {\bibfield  {journal} {\bibinfo
  {journal} {New Journal of Physics}\ }\textbf {\bibinfo {volume} {17}},\
  \bibinfo {pages} {050202} (\bibinfo {year} {2015})}\BibitemShut {NoStop}%
\bibitem [{\citenamefont {Premasiri}\ and\ \citenamefont
  {Gao}(2019)}]{Premasiri_2019}%
  \BibitemOpen
  \bibfield  {author} {\bibinfo {author} {\bibfnamefont {K.}~\bibnamefont
  {Premasiri}}\ and\ \bibinfo {author} {\bibfnamefont {X.~P.~A.}\ \bibnamefont
  {Gao}},\ }\href {\doibase 10.1088/1361-648X/ab04c7} {\bibfield  {journal}
  {\bibinfo  {journal} {Journal of Physics: Condensed Matter}\ }\textbf
  {\bibinfo {volume} {31}},\ \bibinfo {pages} {193001} (\bibinfo {year}
  {2019})}\BibitemShut {NoStop}%
\bibitem [{\citenamefont {Barla}\ \emph {et~al.}(2021)\citenamefont {Barla},
  \citenamefont {Joshi},\ and\ \citenamefont {Bhat}}]{Barla2021}%
  \BibitemOpen
  \bibfield  {author} {\bibinfo {author} {\bibfnamefont {P.}~\bibnamefont
  {Barla}}, \bibinfo {author} {\bibfnamefont {V.~K.}\ \bibnamefont {Joshi}}, \
  and\ \bibinfo {author} {\bibfnamefont {S.}~\bibnamefont {Bhat}},\ }\href
  {\doibase 10.1007/s10825-020-01648-6} {\bibfield  {journal} {\bibinfo
  {journal} {Journal of Computational Electronics}\ }\textbf {\bibinfo {volume}
  {20}},\ \bibinfo {pages} {805} (\bibinfo {year} {2021})}\BibitemShut
  {NoStop}%
\bibitem [{\citenamefont {Guo}\ \emph {et~al.}(2008)\citenamefont {Guo},
  \citenamefont {Murakami}, \citenamefont {Chen},\ and\ \citenamefont
  {Nagaosa}}]{Guo_PhysRevLett.100.096401}%
  \BibitemOpen
  \bibfield  {author} {\bibinfo {author} {\bibfnamefont {G.~Y.}\ \bibnamefont
  {Guo}}, \bibinfo {author} {\bibfnamefont {S.}~\bibnamefont {Murakami}},
  \bibinfo {author} {\bibfnamefont {T.-W.}\ \bibnamefont {Chen}}, \ and\
  \bibinfo {author} {\bibfnamefont {N.}~\bibnamefont {Nagaosa}},\ }\href
  {\doibase 10.1103/PhysRevLett.100.096401} {\bibfield  {journal} {\bibinfo
  {journal} {Phys. Rev. Lett.}\ }\textbf {\bibinfo {volume} {100}},\ \bibinfo
  {pages} {096401} (\bibinfo {year} {2008})}\BibitemShut {NoStop}%
\bibitem [{\citenamefont {Sagasta}\ \emph {et~al.}(2018)\citenamefont
  {Sagasta}, \citenamefont {Omori}, \citenamefont {V\'elez}, \citenamefont
  {Llopis}, \citenamefont {Tollan}, \citenamefont {Chuvilin}, \citenamefont
  {Hueso}, \citenamefont {Gradhand}, \citenamefont {Otani},\ and\ \citenamefont
  {Casanova}}]{Sagasta_PhysRevB.98.060410}%
  \BibitemOpen
  \bibfield  {author} {\bibinfo {author} {\bibfnamefont {E.}~\bibnamefont
  {Sagasta}}, \bibinfo {author} {\bibfnamefont {Y.}~\bibnamefont {Omori}},
  \bibinfo {author} {\bibfnamefont {S.}~\bibnamefont {V\'elez}}, \bibinfo
  {author} {\bibfnamefont {R.}~\bibnamefont {Llopis}}, \bibinfo {author}
  {\bibfnamefont {C.}~\bibnamefont {Tollan}}, \bibinfo {author} {\bibfnamefont
  {A.}~\bibnamefont {Chuvilin}}, \bibinfo {author} {\bibfnamefont {L.~E.}\
  \bibnamefont {Hueso}}, \bibinfo {author} {\bibfnamefont {M.}~\bibnamefont
  {Gradhand}}, \bibinfo {author} {\bibfnamefont {Y.}~\bibnamefont {Otani}}, \
  and\ \bibinfo {author} {\bibfnamefont {F.}~\bibnamefont {Casanova}},\ }\href
  {\doibase 10.1103/PhysRevB.98.060410} {\bibfield  {journal} {\bibinfo
  {journal} {Phys. Rev. B}\ }\textbf {\bibinfo {volume} {98}},\ \bibinfo
  {pages} {060410} (\bibinfo {year} {2018})}\BibitemShut {NoStop}%
\bibitem [{\citenamefont {Momma}\ and\ \citenamefont {Izumi}(2011)}]{VESTA}%
  \BibitemOpen
  \bibfield  {author} {\bibinfo {author} {\bibfnamefont {K.}~\bibnamefont
  {Momma}}\ and\ \bibinfo {author} {\bibfnamefont {F.}~\bibnamefont {Izumi}},\
  }\href@noop {} {\bibfield  {journal} {\bibinfo  {journal} {J. Appl.
  Crystallogr.}\ }\textbf {\bibinfo {volume} {44}},\ \bibinfo {pages} {1272}
  (\bibinfo {year} {2011})}\BibitemShut {NoStop}%
\bibitem [{\citenamefont {Hohenberg}\ and\ \citenamefont
  {Kohn}(1964)}]{HK_dft_1964}%
  \BibitemOpen
  \bibfield  {author} {\bibinfo {author} {\bibfnamefont {P.}~\bibnamefont
  {Hohenberg}}\ and\ \bibinfo {author} {\bibfnamefont {W.}~\bibnamefont
  {Kohn}},\ }\href {\doibase 10.1103/PhysRev.136.B864} {\bibfield  {journal}
  {\bibinfo  {journal} {Phys. Rev.}\ }\textbf {\bibinfo {volume} {136}},\
  \bibinfo {pages} {B864} (\bibinfo {year} {1964})}\BibitemShut {NoStop}%
\bibitem [{\citenamefont {Kohn}\ and\ \citenamefont
  {Sham}(1965)}]{KS_dft_1965}%
  \BibitemOpen
  \bibfield  {author} {\bibinfo {author} {\bibfnamefont {W.}~\bibnamefont
  {Kohn}}\ and\ \bibinfo {author} {\bibfnamefont {L.~J.}\ \bibnamefont
  {Sham}},\ }\href {\doibase 10.1103/PhysRev.140.A1133} {\bibfield  {journal}
  {\bibinfo  {journal} {Phys. Rev.}\ }\textbf {\bibinfo {volume} {140}},\
  \bibinfo {pages} {A1133} (\bibinfo {year} {1965})}\BibitemShut {NoStop}%
\bibitem [{\citenamefont {Kresse}\ and\ \citenamefont
  {Furthm\"uller}(1996{\natexlab{a}})}]{Kresse96a}%
  \BibitemOpen
  \bibfield  {author} {\bibinfo {author} {\bibfnamefont {G.}~\bibnamefont
  {Kresse}}\ and\ \bibinfo {author} {\bibfnamefont {J.}~\bibnamefont
  {Furthm\"uller}},\ }\href@noop {} {\bibfield  {journal} {\bibinfo  {journal}
  {Phys. Rev. B}\ }\textbf {\bibinfo {volume} {54}},\ \bibinfo {pages} {11169}
  (\bibinfo {year} {1996}{\natexlab{a}})}\BibitemShut {NoStop}%
\bibitem [{\citenamefont {Kresse}\ and\ \citenamefont
  {Furthm\"uller}(1996{\natexlab{b}})}]{Kresse96b}%
  \BibitemOpen
  \bibfield  {author} {\bibinfo {author} {\bibfnamefont {G.}~\bibnamefont
  {Kresse}}\ and\ \bibinfo {author} {\bibfnamefont {J.}~\bibnamefont
  {Furthm\"uller}},\ }\href@noop {} {\bibfield  {journal} {\bibinfo  {journal}
  {Comput. Mater. Sci.}\ }\textbf {\bibinfo {volume} {6}},\ \bibinfo {pages}
  {15 } (\bibinfo {year} {1996}{\natexlab{b}})}\BibitemShut {NoStop}%
\bibitem [{\citenamefont {Kresse}\ and\ \citenamefont
  {Joubert}(1999)}]{KressePAW}%
  \BibitemOpen
  \bibfield  {author} {\bibinfo {author} {\bibfnamefont {G.}~\bibnamefont
  {Kresse}}\ and\ \bibinfo {author} {\bibfnamefont {D.}~\bibnamefont
  {Joubert}},\ }\href@noop {} {\bibfield  {journal} {\bibinfo  {journal} {Phys.
  Rev. B}\ }\textbf {\bibinfo {volume} {59}},\ \bibinfo {pages} {1758}
  (\bibinfo {year} {1999})}\BibitemShut {NoStop}%
\bibitem [{\citenamefont {Bl\"ochl}(1994)}]{Blochl94}%
  \BibitemOpen
  \bibfield  {author} {\bibinfo {author} {\bibfnamefont {P.~E.}\ \bibnamefont
  {Bl\"ochl}},\ }\href@noop {} {\bibfield  {journal} {\bibinfo  {journal}
  {Phys. Rev. B}\ }\textbf {\bibinfo {volume} {50}},\ \bibinfo {pages} {17953}
  (\bibinfo {year} {1994})}\BibitemShut {NoStop}%
\bibitem [{\citenamefont {Perdew}\ \emph {et~al.}(2008)\citenamefont {Perdew},
  \citenamefont {Ruzsinszky}, \citenamefont {Csonka}, \citenamefont {Vydrov},
  \citenamefont {Scuseria}, \citenamefont {Constantin}, \citenamefont {Zhou},\
  and\ \citenamefont {Burke}}]{PBEsol}%
  \BibitemOpen
  \bibfield  {author} {\bibinfo {author} {\bibfnamefont {J.~P.}\ \bibnamefont
  {Perdew}}, \bibinfo {author} {\bibfnamefont {A.}~\bibnamefont {Ruzsinszky}},
  \bibinfo {author} {\bibfnamefont {G.~I.}\ \bibnamefont {Csonka}}, \bibinfo
  {author} {\bibfnamefont {O.~A.}\ \bibnamefont {Vydrov}}, \bibinfo {author}
  {\bibfnamefont {G.~E.}\ \bibnamefont {Scuseria}}, \bibinfo {author}
  {\bibfnamefont {L.~A.}\ \bibnamefont {Constantin}}, \bibinfo {author}
  {\bibfnamefont {X.}~\bibnamefont {Zhou}}, \ and\ \bibinfo {author}
  {\bibfnamefont {K.}~\bibnamefont {Burke}},\ }\href@noop {} {\bibfield
  {journal} {\bibinfo  {journal} {Phys. Rev. Lett.}\ }\textbf {\bibinfo
  {volume} {100}},\ \bibinfo {pages} {136406} (\bibinfo {year}
  {2008})}\BibitemShut {NoStop}%
\bibitem [{\citenamefont {Togo}\ and\ \citenamefont {Tanaka}(2015)}]{phonopy}%
  \BibitemOpen
  \bibfield  {author} {\bibinfo {author} {\bibfnamefont {A.}~\bibnamefont
  {Togo}}\ and\ \bibinfo {author} {\bibfnamefont {I.}~\bibnamefont {Tanaka}},\
  }\href@noop {} {\bibfield  {journal} {\bibinfo  {journal} {Scr. Mater.}\
  }\textbf {\bibinfo {volume} {108}},\ \bibinfo {pages} {1} (\bibinfo {year}
  {2015})}\BibitemShut {NoStop}%
\bibitem [{\citenamefont {Singh}\ \emph {et~al.}(2018)\citenamefont {Singh},
  \citenamefont {Valencia-Jaime}, \citenamefont {Pavlic},\ and\ \citenamefont
  {Romero}}]{MechElastic}%
  \BibitemOpen
  \bibfield  {author} {\bibinfo {author} {\bibfnamefont {S.}~\bibnamefont
  {Singh}}, \bibinfo {author} {\bibfnamefont {I.}~\bibnamefont
  {Valencia-Jaime}}, \bibinfo {author} {\bibfnamefont {O.}~\bibnamefont
  {Pavlic}}, \ and\ \bibinfo {author} {\bibfnamefont {A.~H.}\ \bibnamefont
  {Romero}},\ }\href@noop {} {\bibfield  {journal} {\bibinfo  {journal} {Phys.
  Rev. B}\ }\textbf {\bibinfo {volume} {97}},\ \bibinfo {pages} {054108}
  (\bibinfo {year} {2018})}\BibitemShut {NoStop}%
\bibitem [{\citenamefont {Singh}\ \emph {et~al.}(2021)\citenamefont {Singh},
  \citenamefont {Lang}, \citenamefont {Dovale-Farelo}, \citenamefont {Herath},
  \citenamefont {Tavadze}, \citenamefont {Coudert},\ and\ \citenamefont
  {Romero}}]{Mechelastic_Comp_Phys_comm_2021}%
  \BibitemOpen
  \bibfield  {author} {\bibinfo {author} {\bibfnamefont {S.}~\bibnamefont
  {Singh}}, \bibinfo {author} {\bibfnamefont {L.}~\bibnamefont {Lang}},
  \bibinfo {author} {\bibfnamefont {V.}~\bibnamefont {Dovale-Farelo}}, \bibinfo
  {author} {\bibfnamefont {U.}~\bibnamefont {Herath}}, \bibinfo {author}
  {\bibfnamefont {P.}~\bibnamefont {Tavadze}}, \bibinfo {author} {\bibfnamefont
  {F.-X.}\ \bibnamefont {Coudert}}, \ and\ \bibinfo {author} {\bibfnamefont
  {A.~H.}\ \bibnamefont {Romero}},\ }\href {\doibase
  https://doi.org/10.1016/j.cpc.2021.108068} {\bibfield  {journal} {\bibinfo
  {journal} {Computer Physics Communications}\ }\textbf {\bibinfo {volume}
  {267}},\ \bibinfo {pages} {108068} (\bibinfo {year} {2021})}\BibitemShut
  {NoStop}%
\bibitem [{\citenamefont {Wang}\ \emph {et~al.}(2021)\citenamefont {Wang},
  \citenamefont {Xu}, \citenamefont {Liu}, \citenamefont {Tang},\ and\
  \citenamefont {Geng}}]{VASPKIT}%
  \BibitemOpen
  \bibfield  {author} {\bibinfo {author} {\bibfnamefont {V.}~\bibnamefont
  {Wang}}, \bibinfo {author} {\bibfnamefont {N.}~\bibnamefont {Xu}}, \bibinfo
  {author} {\bibfnamefont {J.-C.}\ \bibnamefont {Liu}}, \bibinfo {author}
  {\bibfnamefont {G.}~\bibnamefont {Tang}}, \ and\ \bibinfo {author}
  {\bibfnamefont {W.-T.}\ \bibnamefont {Geng}},\ }\href {\doibase
  https://doi.org/10.1016/j.cpc.2021.108033} {\bibfield  {journal} {\bibinfo
  {journal} {Computer Physics Communications}\ }\textbf {\bibinfo {volume}
  {267}},\ \bibinfo {pages} {108033} (\bibinfo {year} {2021})}\BibitemShut
  {NoStop}%
\bibitem [{\citenamefont {Herath}\ \emph {et~al.}(2020)\citenamefont {Herath},
  \citenamefont {Tavadze}, \citenamefont {He}, \citenamefont {Bousquet},
  \citenamefont {Singh}, \citenamefont {Muñoz},\ and\ \citenamefont
  {Romero}}]{pyprocar}%
  \BibitemOpen
  \bibfield  {author} {\bibinfo {author} {\bibfnamefont {U.}~\bibnamefont
  {Herath}}, \bibinfo {author} {\bibfnamefont {P.}~\bibnamefont {Tavadze}},
  \bibinfo {author} {\bibfnamefont {X.}~\bibnamefont {He}}, \bibinfo {author}
  {\bibfnamefont {E.}~\bibnamefont {Bousquet}}, \bibinfo {author}
  {\bibfnamefont {S.}~\bibnamefont {Singh}}, \bibinfo {author} {\bibfnamefont
  {F.}~\bibnamefont {Muñoz}}, \ and\ \bibinfo {author} {\bibfnamefont {A.~H.}\
  \bibnamefont {Romero}},\ }\href {\doibase
  https://doi.org/10.1016/j.cpc.2019.107080} {\bibfield  {journal} {\bibinfo
  {journal} {Computer Physics Communications}\ }\textbf {\bibinfo {volume}
  {251}},\ \bibinfo {pages} {107080} (\bibinfo {year} {2020})}\BibitemShut
  {NoStop}%
\bibitem [{\citenamefont {Marzari}\ \emph {et~al.}(2012)\citenamefont
  {Marzari}, \citenamefont {Mostofi}, \citenamefont {Yates}, \citenamefont
  {Souza},\ and\ \citenamefont {Vanderbilt}}]{Marzari2012}%
  \BibitemOpen
  \bibfield  {author} {\bibinfo {author} {\bibfnamefont {N.}~\bibnamefont
  {Marzari}}, \bibinfo {author} {\bibfnamefont {A.~A.}\ \bibnamefont
  {Mostofi}}, \bibinfo {author} {\bibfnamefont {J.~R.}\ \bibnamefont {Yates}},
  \bibinfo {author} {\bibfnamefont {I.}~\bibnamefont {Souza}}, \ and\ \bibinfo
  {author} {\bibfnamefont {D.}~\bibnamefont {Vanderbilt}},\ }\href@noop {}
  {\bibfield  {journal} {\bibinfo  {journal} {Rev. Mod. Phys.}\ }\textbf
  {\bibinfo {volume} {84}},\ \bibinfo {pages} {1419} (\bibinfo {year}
  {2012})}\BibitemShut {NoStop}%
\bibitem [{\citenamefont {Wu}\ \emph {et~al.}(2018)\citenamefont {Wu},
  \citenamefont {Zhang}, \citenamefont {Song}, \citenamefont {Troyer},\ and\
  \citenamefont {Soluyanov}}]{WU2018405}%
  \BibitemOpen
  \bibfield  {author} {\bibinfo {author} {\bibfnamefont {Q.}~\bibnamefont
  {Wu}}, \bibinfo {author} {\bibfnamefont {S.}~\bibnamefont {Zhang}}, \bibinfo
  {author} {\bibfnamefont {H.-F.}\ \bibnamefont {Song}}, \bibinfo {author}
  {\bibfnamefont {M.}~\bibnamefont {Troyer}}, \ and\ \bibinfo {author}
  {\bibfnamefont {A.~A.}\ \bibnamefont {Soluyanov}},\ }\href@noop {} {\bibfield
   {journal} {\bibinfo  {journal} {Comput. Phys. Commun.}\ }\textbf {\bibinfo
  {volume} {224}},\ \bibinfo {pages} {405 } (\bibinfo {year}
  {2018})}\BibitemShut {NoStop}%
\bibitem [{\citenamefont {Karmakar}\ \emph {et~al.}(2023)\citenamefont
  {Karmakar}, \citenamefont {Biswas},\ and\ \citenamefont
  {Saha-Dasgupta}}]{Karmakar_PhysRevB.107.075403}%
  \BibitemOpen
  \bibfield  {author} {\bibinfo {author} {\bibfnamefont {S.}~\bibnamefont
  {Karmakar}}, \bibinfo {author} {\bibfnamefont {R.}~\bibnamefont {Biswas}}, \
  and\ \bibinfo {author} {\bibfnamefont {T.}~\bibnamefont {Saha-Dasgupta}},\
  }\href {\doibase 10.1103/PhysRevB.107.075403} {\bibfield  {journal} {\bibinfo
   {journal} {Phys. Rev. B}\ }\textbf {\bibinfo {volume} {107}},\ \bibinfo
  {pages} {075403} (\bibinfo {year} {2023})}\BibitemShut {NoStop}%
\bibitem [{\citenamefont {Royo}\ and\ \citenamefont
  {Stengel}(2021)}]{Stengel_PhysRevX.11.041027}%
  \BibitemOpen
  \bibfield  {author} {\bibinfo {author} {\bibfnamefont {M.}~\bibnamefont
  {Royo}}\ and\ \bibinfo {author} {\bibfnamefont {M.}~\bibnamefont {Stengel}},\
  }\href {\doibase 10.1103/PhysRevX.11.041027} {\bibfield  {journal} {\bibinfo
  {journal} {Phys. Rev. X}\ }\textbf {\bibinfo {volume} {11}},\ \bibinfo
  {pages} {041027} (\bibinfo {year} {2021})}\BibitemShut {NoStop}%
\bibitem [{\citenamefont {Singh}\ and\ \citenamefont
  {Romero}(2017)}]{SinghPRB2017}%
  \BibitemOpen
  \bibfield  {author} {\bibinfo {author} {\bibfnamefont {S.}~\bibnamefont
  {Singh}}\ and\ \bibinfo {author} {\bibfnamefont {A.~H.}\ \bibnamefont
  {Romero}},\ }\href {\doibase 10.1103/PhysRevB.95.165444} {\bibfield
  {journal} {\bibinfo  {journal} {Phys. Rev. B}\ }\textbf {\bibinfo {volume}
  {95}},\ \bibinfo {pages} {165444} (\bibinfo {year} {2017})}\BibitemShut
  {NoStop}%
\bibitem [{\citenamefont {Singh}\ \emph {et~al.}(2020)\citenamefont {Singh},
  \citenamefont {Kim}, \citenamefont {Rabe},\ and\ \citenamefont
  {Vanderbilt}}]{SinghPRL2020}%
  \BibitemOpen
  \bibfield  {author} {\bibinfo {author} {\bibfnamefont {S.}~\bibnamefont
  {Singh}}, \bibinfo {author} {\bibfnamefont {J.}~\bibnamefont {Kim}}, \bibinfo
  {author} {\bibfnamefont {K.~M.}\ \bibnamefont {Rabe}}, \ and\ \bibinfo
  {author} {\bibfnamefont {D.}~\bibnamefont {Vanderbilt}},\ }\href {\doibase
  10.1103/PhysRevLett.125.046402} {\bibfield  {journal} {\bibinfo  {journal}
  {Phys. Rev. Lett.}\ }\textbf {\bibinfo {volume} {125}},\ \bibinfo {pages}
  {046402} (\bibinfo {year} {2020})}\BibitemShut {NoStop}%
\bibitem [{\citenamefont {Born}\ \emph {et~al.}(1955)\citenamefont {Born},
  \citenamefont {Huang},\ and\ \citenamefont {Lax}}]{Born_Huang}%
  \BibitemOpen
  \bibfield  {author} {\bibinfo {author} {\bibfnamefont {M.}~\bibnamefont
  {Born}}, \bibinfo {author} {\bibfnamefont {K.}~\bibnamefont {Huang}}, \ and\
  \bibinfo {author} {\bibfnamefont {M.}~\bibnamefont {Lax}},\ }\href {\doibase
  10.1119/1.1934059} {\bibfield  {journal} {\bibinfo  {journal} {American
  Journal of Physics}\ }\textbf {\bibinfo {volume} {23}},\ \bibinfo {pages}
  {474} (\bibinfo {year} {1955})},\ \Eprint
  {http://arxiv.org/abs/https://pubs.aip.org/aapt/ajp/article-pdf/23/7/474/12115493/474\_1\_online.pdf}
  {https://pubs.aip.org/aapt/ajp/article-pdf/23/7/474/12115493/474\_1\_online.pdf}
  \BibitemShut {NoStop}%
\bibitem [{\citenamefont {Nye}(1985)}]{nye1985physical}%
  \BibitemOpen
  \bibfield  {author} {\bibinfo {author} {\bibfnamefont {J.~F.}\ \bibnamefont
  {Nye}},\ }\href@noop {} {\emph {\bibinfo {title} {Physical properties of
  crystals: their representation by tensors and matrices}}}\ (\bibinfo
  {publisher} {Oxford university press},\ \bibinfo {year} {1985})\BibitemShut
  {NoStop}%
\bibitem [{\citenamefont {Andrew}\ \emph {et~al.}(2012)\citenamefont {Andrew},
  \citenamefont {Mapasha}, \citenamefont {Ukpong},\ and\ \citenamefont
  {Chetty}}]{Andrew_PhysRevB.85.125428}%
  \BibitemOpen
  \bibfield  {author} {\bibinfo {author} {\bibfnamefont {R.~C.}\ \bibnamefont
  {Andrew}}, \bibinfo {author} {\bibfnamefont {R.~E.}\ \bibnamefont {Mapasha}},
  \bibinfo {author} {\bibfnamefont {A.~M.}\ \bibnamefont {Ukpong}}, \ and\
  \bibinfo {author} {\bibfnamefont {N.}~\bibnamefont {Chetty}},\ }\href
  {\doibase 10.1103/PhysRevB.85.125428} {\bibfield  {journal} {\bibinfo
  {journal} {Phys. Rev. B}\ }\textbf {\bibinfo {volume} {85}},\ \bibinfo
  {pages} {125428} (\bibinfo {year} {2012})}\BibitemShut {NoStop}%
\bibitem [{\citenamefont {Vajna}\ \emph {et~al.}(2012)\citenamefont {Vajna},
  \citenamefont {Simon}, \citenamefont {Szilva}, \citenamefont {Palotas},
  \citenamefont {Ujfalussy},\ and\ \citenamefont
  {Szunyogh}}]{Vajna_PhysRevB.85.075404}%
  \BibitemOpen
  \bibfield  {author} {\bibinfo {author} {\bibfnamefont {S.}~\bibnamefont
  {Vajna}}, \bibinfo {author} {\bibfnamefont {E.}~\bibnamefont {Simon}},
  \bibinfo {author} {\bibfnamefont {A.}~\bibnamefont {Szilva}}, \bibinfo
  {author} {\bibfnamefont {K.}~\bibnamefont {Palotas}}, \bibinfo {author}
  {\bibfnamefont {B.}~\bibnamefont {Ujfalussy}}, \ and\ \bibinfo {author}
  {\bibfnamefont {L.}~\bibnamefont {Szunyogh}},\ }\href {\doibase
  10.1103/PhysRevB.85.075404} {\bibfield  {journal} {\bibinfo  {journal} {Phys.
  Rev. B}\ }\textbf {\bibinfo {volume} {85}},\ \bibinfo {pages} {075404}
  (\bibinfo {year} {2012})}\BibitemShut {NoStop}%
\bibitem [{\citenamefont {Yang}\ and\ \citenamefont
  {Chang}(2006)}]{Yang_PhysRevB.74.193314}%
  \BibitemOpen
  \bibfield  {author} {\bibinfo {author} {\bibfnamefont {W.}~\bibnamefont
  {Yang}}\ and\ \bibinfo {author} {\bibfnamefont {K.}~\bibnamefont {Chang}},\
  }\href {\doibase 10.1103/PhysRevB.74.193314} {\bibfield  {journal} {\bibinfo
  {journal} {Phys. Rev. B}\ }\textbf {\bibinfo {volume} {74}},\ \bibinfo
  {pages} {193314} (\bibinfo {year} {2006})}\BibitemShut {NoStop}%
\bibitem [{\citenamefont {Gupta}\ and\ \citenamefont
  {Yakobson}(2021)}]{Gupta_JACS_2021}%
  \BibitemOpen
  \bibfield  {author} {\bibinfo {author} {\bibfnamefont {S.}~\bibnamefont
  {Gupta}}\ and\ \bibinfo {author} {\bibfnamefont {B.~I.}\ \bibnamefont
  {Yakobson}},\ }\href {\doibase 10.1021/jacs.0c12809} {\bibfield  {journal}
  {\bibinfo  {journal} {Journal of the American Chemical Society}\ }\textbf
  {\bibinfo {volume} {143}},\ \bibinfo {pages} {3503} (\bibinfo {year}
  {2021})}\BibitemShut {NoStop}%
\bibitem [{\citenamefont {Soluyanov}\ and\ \citenamefont
  {Vanderbilt}(2011)}]{Soluyanov_PhysRevB.83.035108}%
  \BibitemOpen
  \bibfield  {author} {\bibinfo {author} {\bibfnamefont {A.~A.}\ \bibnamefont
  {Soluyanov}}\ and\ \bibinfo {author} {\bibfnamefont {D.}~\bibnamefont
  {Vanderbilt}},\ }\href {\doibase 10.1103/PhysRevB.83.035108} {\bibfield
  {journal} {\bibinfo  {journal} {Phys. Rev. B}\ }\textbf {\bibinfo {volume}
  {83}},\ \bibinfo {pages} {035108} (\bibinfo {year} {2011})}\BibitemShut
  {NoStop}%
\bibitem [{\citenamefont {Hirsch}(1999)}]{Hirsch_PhysRevLett.83.1834}%
  \BibitemOpen
  \bibfield  {author} {\bibinfo {author} {\bibfnamefont {J.~E.}\ \bibnamefont
  {Hirsch}},\ }\href {\doibase 10.1103/PhysRevLett.83.1834} {\bibfield
  {journal} {\bibinfo  {journal} {Phys. Rev. Lett.}\ }\textbf {\bibinfo
  {volume} {83}},\ \bibinfo {pages} {1834} (\bibinfo {year}
  {1999})}\BibitemShut {NoStop}%
\bibitem [{\citenamefont {Dyakonov}\ and\ \citenamefont
  {Perel}(1971)}]{DYAKONOV_phy_let_a_1971459}%
  \BibitemOpen
  \bibfield  {author} {\bibinfo {author} {\bibfnamefont {M.}~\bibnamefont
  {Dyakonov}}\ and\ \bibinfo {author} {\bibfnamefont {V.}~\bibnamefont
  {Perel}},\ }\href {\doibase https://doi.org/10.1016/0375-9601(71)90196-4}
  {\bibfield  {journal} {\bibinfo  {journal} {Physics Letters A}\ }\textbf
  {\bibinfo {volume} {35}},\ \bibinfo {pages} {459} (\bibinfo {year}
  {1971})}\BibitemShut {NoStop}%
\bibitem [{\citenamefont {Jungwirth}\ \emph {et~al.}(2012)\citenamefont
  {Jungwirth}, \citenamefont {Wunderlich},\ and\ \citenamefont
  {Olejn{\'i}k}}]{Jungwirth_Nature_Mat_2012}%
  \BibitemOpen
  \bibfield  {author} {\bibinfo {author} {\bibfnamefont {T.}~\bibnamefont
  {Jungwirth}}, \bibinfo {author} {\bibfnamefont {J.}~\bibnamefont
  {Wunderlich}}, \ and\ \bibinfo {author} {\bibfnamefont {K.}~\bibnamefont
  {Olejn{\'i}k}},\ }\href {\doibase 10.1038/nmat3279} {\bibfield  {journal}
  {\bibinfo  {journal} {Nature Materials}\ }\textbf {\bibinfo {volume} {11}},\
  \bibinfo {pages} {382} (\bibinfo {year} {2012})}\BibitemShut {NoStop}%
\bibitem [{\citenamefont {Gong}\ \emph {et~al.}(2024)\citenamefont {Gong},
  \citenamefont {Li}, \citenamefont {Wang},\ and\ \citenamefont
  {Zhang}}]{Gong_PhysRevB.109.045124}%
  \BibitemOpen
  \bibfield  {author} {\bibinfo {author} {\bibfnamefont {L.}~\bibnamefont
  {Gong}}, \bibinfo {author} {\bibfnamefont {Y.}~\bibnamefont {Li}}, \bibinfo
  {author} {\bibfnamefont {H.}~\bibnamefont {Wang}}, \ and\ \bibinfo {author}
  {\bibfnamefont {H.}~\bibnamefont {Zhang}},\ }\href {\doibase
  10.1103/PhysRevB.109.045124} {\bibfield  {journal} {\bibinfo  {journal}
  {Phys. Rev. B}\ }\textbf {\bibinfo {volume} {109}},\ \bibinfo {pages}
  {045124} (\bibinfo {year} {2024})}\BibitemShut {NoStop}%
\bibitem [{\citenamefont {Farzaneh}\ and\ \citenamefont
  {Rakheja}(2020)}]{Farzaneh_PhysRevMaterials.4.114202}%
  \BibitemOpen
  \bibfield  {author} {\bibinfo {author} {\bibfnamefont {S.~M.}\ \bibnamefont
  {Farzaneh}}\ and\ \bibinfo {author} {\bibfnamefont {S.}~\bibnamefont
  {Rakheja}},\ }\href {\doibase 10.1103/PhysRevMaterials.4.114202} {\bibfield
  {journal} {\bibinfo  {journal} {Phys. Rev. Mater.}\ }\textbf {\bibinfo
  {volume} {4}},\ \bibinfo {pages} {114202} (\bibinfo {year}
  {2020})}\BibitemShut {NoStop}%
\bibitem [{\citenamefont {Gao}\ \emph {et~al.}(2021)\citenamefont {Gao},
  \citenamefont {Wu}, \citenamefont {Persson},\ and\ \citenamefont
  {Wang}}]{Bilbao}%
  \BibitemOpen
  \bibfield  {author} {\bibinfo {author} {\bibfnamefont {J.}~\bibnamefont
  {Gao}}, \bibinfo {author} {\bibfnamefont {Q.}~\bibnamefont {Wu}}, \bibinfo
  {author} {\bibfnamefont {C.}~\bibnamefont {Persson}}, \ and\ \bibinfo
  {author} {\bibfnamefont {Z.}~\bibnamefont {Wang}},\ }\href {\doibase
  https://doi.org/10.1016/j.cpc.2020.107760} {\bibfield  {journal} {\bibinfo
  {journal} {Computer Physics Communications}\ }\textbf {\bibinfo {volume}
  {261}},\ \bibinfo {pages} {107760} (\bibinfo {year} {2021})}\BibitemShut
  {NoStop}%
\bibitem [{\citenamefont {Wang}\ \emph {et~al.}(2020)\citenamefont {Wang},
  \citenamefont {Gopal}, \citenamefont {Picozzi}, \citenamefont {Curtarolo},
  \citenamefont {Buongiorno~Nardelli},\ and\ \citenamefont
  {S{\l}awi{\'{n}}ska}}]{Wang_npj_2020}%
  \BibitemOpen
  \bibfield  {author} {\bibinfo {author} {\bibfnamefont {H.}~\bibnamefont
  {Wang}}, \bibinfo {author} {\bibfnamefont {P.}~\bibnamefont {Gopal}},
  \bibinfo {author} {\bibfnamefont {S.}~\bibnamefont {Picozzi}}, \bibinfo
  {author} {\bibfnamefont {S.}~\bibnamefont {Curtarolo}}, \bibinfo {author}
  {\bibfnamefont {M.}~\bibnamefont {Buongiorno~Nardelli}}, \ and\ \bibinfo
  {author} {\bibfnamefont {J.}~\bibnamefont {S{\l}awi{\'{n}}ska}},\ }\href
  {\doibase 10.1038/s41524-020-0274-0} {\bibfield  {journal} {\bibinfo
  {journal} {npj Computational Materials}\ }\textbf {\bibinfo {volume} {6}},\
  \bibinfo {pages} {7} (\bibinfo {year} {2020})}\BibitemShut {NoStop}%
\bibitem [{\citenamefont {Zuo}\ \emph {et~al.}(2023)\citenamefont {Zuo},
  \citenamefont {Feng}, \citenamefont {Liu}, \citenamefont {Huang},
  \citenamefont {Liu}, \citenamefont {Liu},\ and\ \citenamefont
  {Cui}}]{Zuo_PhysRevB.108.195129}%
  \BibitemOpen
  \bibfield  {author} {\bibinfo {author} {\bibfnamefont {X.}~\bibnamefont
  {Zuo}}, \bibinfo {author} {\bibfnamefont {Y.}~\bibnamefont {Feng}}, \bibinfo
  {author} {\bibfnamefont {N.}~\bibnamefont {Liu}}, \bibinfo {author}
  {\bibfnamefont {B.}~\bibnamefont {Huang}}, \bibinfo {author} {\bibfnamefont
  {M.}~\bibnamefont {Liu}}, \bibinfo {author} {\bibfnamefont {D.}~\bibnamefont
  {Liu}}, \ and\ \bibinfo {author} {\bibfnamefont {B.}~\bibnamefont {Cui}},\
  }\href {\doibase 10.1103/PhysRevB.108.195129} {\bibfield  {journal} {\bibinfo
   {journal} {Phys. Rev. B}\ }\textbf {\bibinfo {volume} {108}},\ \bibinfo
  {pages} {195129} (\bibinfo {year} {2023})}\BibitemShut {NoStop}%
\end{thebibliography}%
\end{document}